\documentclass{article}

\usepackage{arxiv}

\usepackage[utf8]{inputenc} % allow utf-8 input
\usepackage[T1]{fontenc}    % use 8-bit T1 fonts
\usepackage{hyperref}       % hyperlinks
\usepackage{url}            % simple URL typesetting
\usepackage{booktabs}       % professional-quality tables
\usepackage{amsfonts}       % blackboard math symbols
\usepackage{nicefrac}       % compact symbols for 1/2, etc.
\usepackage{microtype}      % microtypography
\usepackage{lipsum}
\usepackage{slashed}

\usepackage{graphicx}
\usepackage{xcolor}
\usepackage{amsmath}
\usepackage{tikz}
\usepackage{pgfplots}
\usepackage{multicol}
\usepackage{multirow}
\usepackage{subfig}
\usepackage{comment}
\usepackage{appendix}
\DeclareMathOperator{\sgn}{sgn}

\title{A comparison between the convected level-set and the Allen-Cahn phase-field methods for the simulation of two-phase flows}
\title{Comparing the convected level-set and the Allen-Cahn phase-field methods in AMR/C simulations of two-phase flows}

\author{
  Malú Grave \\
  Dept. of Civil Engineering\\
  COPPE/Federal University of Rio de Janeiro\\
  P.O. Box 68506, RJ 21945-970, Rio de Janeiro, Brazil \\
  Fundação Oswaldo Cruz – Fiocruz \\
  Rua Waldemar Falcão, 121, Candeal, BA 40296-710, Salvador, Brazil \\
  \texttt{malugrave@nacad.ufrj.br} \\
   \And
 Alvaro L.G.A. Coutinho \\
  Dept. of Civil Engineering\\
  COPPE/Federal University of Rio de Janeiro \\
  P.O. Box 68506, RJ 21945-970, Rio de Janeiro, Brazil \\
  \texttt{alvaro@nacad.ufrj.br} \\
}

\begin{document}
\maketitle

\begin{abstract}
The modeling and simulation of two-phase flows is still an active research area, mainly when surface tension is present. One way to model the different phases is with interface capturing methods. Two well-established interface capturing approaches are the level-set and phase-field methods. The level-set method is known for its ability to compute the surface tension accurately, and phase-field models satisfy the second law of thermodynamics. This paper reviews and compares a level-set and a phase-field approach to simulate two-phase flows. 
We use a modified level-set method, called convected level-set. The difference to the standard level-set method is that the re-initialization step is embedded in the convection equation, avoiding a separate step during the calculation. We also apply a global mass conservation procedure to enforce the mass balance between phases. On the other hand, the phase-field approach uses a conservative Allen-Cahn equation with a Lagrange multiplier to conserve the mass. The methods are implemented in \texttt{libMesh}, a parallel adaptive finite element library, using the same finite element formulations, time-marching schemes, solvers, and mesh adaptivity strategies. We present numerical solutions for the two methods considering adaptive mesh refinement and coarsening (AMR/C). Results are analyzed and discussed.
\end{abstract}

% keywords can be removed

\section{Introduction}

The modeling and simulation of two-phase flows is still an area of active research. Many relevant industrial problems involve two-phase fluid flows. Two cases of interest are the droplet impact on a solid surface and rising bubble dynamics. The drop impact on a solid or liquid surface is a common phenomenon that occurs in many situations. These include industrial applications such as welding in material processing, coating, painting, cooling, fuel injection in internal combustion engines; also ink-jet printing, agricultural aspects related to the rainfall as soil erosion, among others \cite{chinakhov2013calculation, moreira2010advances, derby2011inkjet, joung2015aerosol}. Understanding the physics of drop impact on wet walls is essential in optimizing these applications. The study of bubbles' motion is of fundamental importance in many physical, chemical, and biological processes such as boiling, cloud cavitation in hydraulic systems, bubble columns and centrifuges in the petrochemical industry, bubble nucleation and cooperativity in DNA melting, fuel atomization, and chemical reactions in the combustion process \cite{khalloufi2019adaptive,luo2016review,besagni2018two,ares2005bubble,desjardins2008accurate}. The study of bubbles is also essential to understand natural phenomena such as the sound propagation in the ocean, the exchange of gases and heat between the oceans and the atmosphere, and explosive volcanic eruptions \cite{BENILOV2015144,lyons2019infrasound}. Several experimental studies on the bubbles and droplets dynamics have been conducted. Although experiments provide reliable results, they are difficult to reproduce, and measuring all quantities of interest can be challenging. Thus, numerical simulation has become an alternative approach for such complicated studies.

In numerical simulations, one of the main issues is to model the motion and deformation of the interface between the two phases. Two candidates for modeling bubbles motion and deformation are the level-set  \cite{Grave_Camata_Coutinho_2020} and the phase-field methods \cite{khanwale2019simulating}. In this work, we focus on reviewing and comparing these two methods.

In the case of the level-set method, the interface is represented as the zero contour of a signed distance function, in which positive values represent one phase while negatives the other one. A convection equation governs the movement of the interface. After the convection, there is no guarantee that the level-set function keeps its signed distance function properties. Therefore, a re-initialization process is needed. One of the drawbacks of the level-set method is that discrete solutions of the level-set equation do not generally conserve mass between phases. While the mass errors may be negligible for highly resolved simulations or short simulations, long-time simulations or coarse grid simulations can fail catastrophically, mainly because these errors may produce incorrect fluid field distributions. Several attempts to improve mass conservation of the level-set method have been made. In \cite{sussman2000coupled, tsui2017coupled, cao2019coupled, wang2009coupled} a combination of the level-set method and the Volume of Fluid (VOF) method was used to obtain the mass conservation of the VOF method while using a level-set function to obtain better approximations of the surface tension. In the conservative level-set (CLS) method \cite{olsson2005conservative, olsson2007conservative, quezada2020unstructured}, an approximate Heaviside function is used instead of the standard level-set function so that the conserved quantity in the convection equation approximates the mass of one of the phases. Some authors have combined the CLS method with a finite-volume spatial discretization to circumvent the accumulation of mass conservation error \cite{balcazar2014finite, balcazar2019level}. Others have taken advantage of the fact that the mass conservation error reduces when using adaptive mesh refinement by refining the mesh in the regions close to the interface \cite{min2007second, gibou2018review}. Other methods consist essentially in a post-processing step for the standard level-set approach, in which mass conservation is achieved by adjusting the level-set function and consequently, the mass between phases \cite{kees2011conservative, de2019monolithic, de2019unstructured}. Another issue is that the level-set formulations, different from the phase-field formulations, are not energy-stable. Therefore, a model that lies in between phase-field and level-set formulations was developed \cite{ten2021novel}. Also, the first energy-dissipative level-set method for the incompressible Navier-Stokes equations with surface tension is presented in \cite{eikelder2020energy}. In \cite{khalloufi2016high} the authors show the importance of a dynamic unstructured anisotropic mesh adaptation to represent better the surface tension and change of phases; however, this technique will not be considered here.

In the present paper, we assess the convected level-set method as presented in \cite{Grave_Camata_Coutinho_2020}. This method associates both re-initialization and convection steps. The re-initialization inclusion in the convection equation avoids the extra step that appears in the original level-set formulation. We use a truncated signed distance function to get a smooth transition close to the interface and improve the mass conservation. This signed distance function follows the idea of the CLS methods of replacing the signed distance function of the standard level-set method for a truncated function near the interface. This modification significantly improves mass conservation. However,  we add a further step to guarantee this condition. If there is a mass gain or loss, the mass conservation between phases is enforced by a global mass conservation procedure based on \cite{smolianski2001numerical}.

In the case of the phase-field method, sharp fluid interfaces are replaced by thin but nonzero thickness transition regions where the interfacial forces are smoothly distributed. Phase-field
models also satisfy the second law of thermodynamics. Two fundamental equations of this method are the Allen-Cahn \cite{allen1979microscopic}, and Cahn-Hilliard \cite{cahn1958free} equations, which are originally introduced to describe the non-conservative and conservative phase variables in the phase separation process, respectively. The Cahn-Hilliard equation conserves mass between phases but has the drawback of being a fourth-order differential equation, which implementation is not trivial. On the other hand, the Allen-Cahn equation is simpler to implement but has the drawback of not conserving mass. Therefore, conservative Allen-Cahn equations have arisen. Usually, Allen-Cahn equations are implemented with a time-dependent Lagrange multiplier to enforce conservation of mass \cite{kim2014conservative}. Thus, mass conservation is guaranteed, and the implementation does not require methods for high-order differential equations. The phase-field equations also make use of a mobility coefficient. Many studies have assumed that mobility is constant. In \cite{vasconcelos2014residual} the mobility coefficient was evaluated exploring the Allen-Cahn equation similarities with residual-based discontinuity-capturing schemes, making the equation dependent on its residual. In general, numerical methods and simulations with variable mobility have shown that a time-dependent mobility coefficient captures the interface motion in a better way \cite{mao2021variational, zhu1999coarsening, wells2006discontinuous,shin2019cahn}. Phase-field models are extensively used in multi-phase problems \cite{aihara2019multi, yang2022fully}, fracture mechanics \cite{wu2020phase,zhou2018phase} and also may be applied to fluid-structure interaction problems \cite{rath2021adaptive}.

In this work, we use an Allen-Cahn equation as described in \cite{mao2021variational}. In this formulation, the mobility coefficient is adjusted adaptively as a function of gradients of the velocity and the order parameter in the diffuse interface region in such a way that the free energy minimization correctly opposes the convective distortion. The mass conservation is achieved by enforcing a Lagrange multiplier with temporal and spatial dependence on the Allen-Cahn phase-field function.

Both interface capturing methods, discretized by stabilized finite formulations, are coupled with the Navier-Stokes equations that are treated with the residual-based variational multiscale finite element formulation. All implementations in this work are done using the \texttt{libMesh} library. \texttt{libMesh} is an open-source library that provides a platform for parallel, adaptive, multiphysics finite element simulations \cite{libmesh}. The main advantage of \texttt{libMesh} is the possibility to focus on the implementation of modeling specific features
without worrying about issues such as adaptivity and code parallelization. Consequently, the effort to build a high-performance computing code tends to be minimized. Moreover, we consider adaptive mesh refinement and coarsening (AMR/C) based on the flux jump of the phase variable, the order parameter, in our simulations.

To summarize, the objective of this study is to compare the convected level-set and the Allen-Cahn phase-field methods applied to two-phase flow problems considering adaptive mesh refinement and coarsening. The remainder of this paper is organized as follows. We start in section \ref{PF} with a description of the Allen-Cahn phase-field governing equations. Then, we present the convected level-set method. In section \ref{NSE}, we present the Navier-Stokes equations for two-phase flows. Section \ref{AMR} describes the implementation details briefly. Finally, section \ref{Results} provides the validation of the implementation of the models and the simulations obtained from both methods. The paper ends with a summary of our main conclusions.

\section{Allen-Cahn phase-field method}\label{PF}

The Allen-Cahn phase-field method is used to model the interface between two phases, which considers a diffuse representation of the interface geometry and
describes the minimization of the free energy functional \cite{cahn1958free}. The diffuse interface between the two phases is described as a region where the phases are mixed and store the free energy. We employ the Allen-Cahn phase-field equation with a Lagrange multiplier
for solving two-phase flow problems in the current study by its computational efficiency and stability. Therefore, the motion of the phase-field is described by  \cite{mao2021variational}:

\begin{equation}\label{phasefield1}
\begin{split}
\frac{\partial \phi}{\partial t} + \mathbf{u}\cdot \nabla \phi - \gamma(t) \bigg(\epsilon^2 \nabla^2\phi - F'(\phi) + \beta(t) \sqrt{F(\phi)}\bigg) = 0  \textrm{ in }\Omega \times [0,t_f]
\\
\nabla \phi = 0 \textrm{ in }\Gamma \times [0,t_f]
\\
\phi(\mathbf{x},0) = \phi_0(\mathbf{x})
\end{split}
\end{equation} 

\noindent where $\phi(\mathbf{x},t) \in \Omega \subset R^{nsd}$, with boundary $\Gamma \in R^{nsd -1}$ represents the mixture of the phases (pure phases are $\phi$ = 1 and $\phi$ = -1), $\mathbf{u}$ is the velocity field, $\epsilon$ is the thickness of the diffuse interface layer,  $\gamma(t)$ is a time-dependent mobility coefficient, given by,

\begin{equation}\label{gamma}
\gamma(t) = \frac{1}{\eta} \mathcal{F}\left( \left|\left| \frac{\nabla \phi \cdot \nabla \mathbf{u} \cdotp \nabla \phi}{||\nabla \phi||^2} \right|\right| \right)
\end{equation} 

\noindent where $\mathcal{F}(\psi(\mathbf{x},t)) = \sqrt{\frac{\int_\Omega (\psi(\mathbf{x},t))^2d\Omega}{V_\Omega}}$, $V_\Omega$ is the volume (or area) of the domain, and $\eta$ is the RMS convective distortion parameter. 
The term $F'(\phi)$
denotes the derivative of $F(\phi)$ with respect to $\phi$, being $F(\phi)$ the double-well energy potential $F(\phi) = \frac{1}{4}(\phi^2-1)^2$. The parameter $\beta(t)$ is the time dependent part of the Lagrange multiplier, given by,

\begin{equation}\label{beta}
\beta(t) = \frac{\int_\Omega F'(\phi)d\Omega}{\int_\Omega \sqrt{F(\phi)}d\Omega}
\end{equation} 

For the initial conditions $\phi_0(\mathbf{x})$, the equation is:

\begin{equation}\label{phasefield2}
\phi(d(\mathbf{x},\Gamma)) = \tanh\left(\frac{d(\mathbf{x},\Gamma)}{\sqrt{2}\epsilon}\right)
\end{equation} 

\noindent where $d(\mathbf{x},\Gamma)$ is the Euclidian distance to the interface $\Gamma$.

The implementation is done with a SUPG finite element variational formulation \cite{brooks1982streamline} and for time integration, we use the second-order Backward Differentiation Formula (BDF2).

\section{Convected level-set Method}\label{CLS}

The level-set method was first introduced by \cite{osher1988} in the late 1980s as a technique for capturing evolving interfaces and tracking the propagation of fronts. The method consists of separating two phases with signed distance functions (SDFs), in which the zero level-set defines the interface between phases. In this work, we use the convected level-set method as shown in \cite{Grave_Camata_Coutinho_2020}. Following \cite{Grave_Camata_Coutinho_2020}, we use a modified SDF ($\phi$), inspired in \cite{bahbah2019conservative}, given by,

\begin{equation}
\phi = \frac{1}{1+e^{\frac{-\alpha}{E}}} -0.5
\label{lsfunction}
\end{equation} 

\noindent in which the parameter $E$ defines the thickness where the modified SDF is distributed and $\alpha$ is the standard level-set function given by,

\begin{equation}
\begin{split}
&\alpha(\mathbf{x}) = \left\{\begin{array}{ccc}
d(\mathbf{x},\Gamma) \textnormal{ for } \mathbf{x} \in \Omega_{1}
\\
0 \textnormal{ for } \mathbf{x} \in \Gamma
\\
-d(\mathbf{x},\Gamma) \textnormal{ for } \mathbf{x} \in \Omega_{2} .
\end{array}\right.
\end{split}
\label{signed_function}
\end{equation}

\noindent where $\Omega_1$ and $\Omega_2$ are the different phases of the fluid in $\Omega$.

%In this work, we define $E=2h_e$, with $h_e$ being the mesh size.
The bigger the value of $E$, the smoother is the transition between phases. However, as in the CLS methods \cite{olsson2005conservative, olsson2007conservative}, we would like a sharp transition to improve mass conservation. %the bigger is the region where there will be iso-values of the level-set function. 
The idea is to minimize this region as much as possible and still have a smooth transition between phases. The truncated level-set function verifies the following property,

\begin{equation}
S=||\nabla \phi||= \frac{1}{4E} - \frac{\phi^2}{E}.
\end{equation}

The convected level-set method avoids the re-initialization step of the original formulation by combining the re-initialization with the convection equation in one single equation \cite{coupez2007convection,ville2011,toure2016stabilized,Grave_Camata_Coutinho_2020}. Thus, the interface motion is given by the following equation,

\begin{equation}
\begin{split}
\frac{\partial \phi}{\partial t} + (\mathbf{u} + \lambda \mathbf{U})\cdot \nabla \phi - \lambda \sgn(\phi) \left(\frac{1}{4E} - \frac{\phi^2}{E}\right) 
%- \nabla \cdot (\kappa_{dc}\nabla \phi)
= 0  \textrm{ in }\Omega \times [0,t_f]
\\
\nabla \phi = 0  \textrm{ in }\Gamma \times [0,t_f]
\\
\phi(\mathbf{x},0) = \phi_0(\mathbf{x})
\end{split}
\label{convec_reinit3}
\end{equation}

\noindent where $\mathbf{U} = \sgn(\phi) \dfrac{\nabla \phi}{||\nabla \phi||}$, $\lambda$ is a penalty constant and $\sgn$ is the sign function.

The penalty constant $\lambda$ defines the contribution of the re-initialization equation in the convection equation. A small $\lambda$ may not be enough to correct the iso-surfaces and adequately recover the signed distance properties, while a large one may change the interface shape.

Due to the incompressibility assumption, the volume (area) of the region occupied by each fluid must be conserved during the whole computational process. However, the convected level-set method cannot guarantee mass conservation. Numerical errors may decrease/increase the area/volume of one of the fluids by about several percent after many time steps. That is why some authors combine the level-set method with some enforcement of mass balance (see \cite{chang1996level}, and  \cite{sussman1999efficient}).

Mass conservation means that the current area/volume of one of the phases $V_{\Omega_2}$ is equal to its initial area/volume $V_{0\Omega_2}$. The initial conditions define  $V_{0\Omega_2}$. If $V_{\Omega_2}$ is different from $V_{0\Omega_2}$, we need to introduce a correction. It is necessary only to correct one of the domains since their union is constant. However, we do not want to change the shape of the interface between the two phases. Therefore, we change the zero level-set, accepting as a new zero level-set some near isoline, since it has almost the same shape. To do that, we move the level-set function upward or downward, by adding to $\phi$ a global constant perturbation $c_\phi$, introduced for 2D simulations in \cite{smolianski2001numerical}, and generalized here for any number of spatial dimensions as,

\begin{equation}
c_\phi = \frac{V_{0\Omega_2} - V_{\Omega_2}}{L_t(\Gamma)}
\end{equation}

\noindent where $L_t(\Gamma)$ is the length/area of the interface $\Gamma$.

We can approximate $V_{\Omega_2}$ and $L_t(\Gamma)$ by,

\begin{equation}
V_{\Omega_2}=\int_\Omega \mathcal{H}_\epsilon(\alpha)d_\Omega
\end{equation}

\begin{equation}\label{perimeter-area}
L_t=\int_\Omega \delta_\epsilon(\alpha)d_\Omega
\end{equation}

\noindent being $\mathcal{H}_\epsilon$ the regularized
Heaviside function and $\delta_\epsilon$ the regularized Dirac function.  The functions $\mathcal{H}_\epsilon$ and $\delta_\epsilon$ are,

\begin{equation}
\begin{split}
\mathcal{H}_{\epsilon}(\alpha)=  \left\{\begin{array}{ccc}
0 \textnormal{ for }  \alpha < -\epsilon_{LS} \\
\frac{1}{2}\left(1+\frac{\alpha}{\epsilon_{LS}}+ \frac{1}{\pi}\sin \left(\frac{\pi\alpha}{\epsilon_{LS}}\right)\right)    \textnormal{ for }  |\alpha| \leq \epsilon_{LS}  \\
1  \textnormal{ for }  \alpha > \epsilon_{LS} \end{array}\right.
\end{split}
\end{equation}

 \begin{equation}
\begin{split}
\delta_{\epsilon}(\alpha)=  \left\{\begin{array}{ccc}
0 \textnormal{ for }  \alpha > \epsilon_{LS} \\
\frac{1}{2\epsilon_{LS}}\left(1 + \cos \left(\frac{\pi\alpha}{\epsilon_{LS}}\right) \right)    \textnormal{ for }  |\alpha| \leq \epsilon_{LS} \\
0  \textnormal{ for }  \alpha < \epsilon_{LS}. \end{array}\right.
\end{split}
\end{equation}

\noindent in which $\epsilon_{LS}$ is a thickness related to the mesh size.  Here, $\epsilon_{LS}$ is defined as $\epsilon_{LS}=h_e$, with $h_e$ being the minimal mesh size.

The implementation is done with the SUPG finite element variational formulation and supplemented by a small stabilizing diffusion term, based on the $YZ\beta$ discontinuity-capturing operator \cite{bazilevs}, to improve stability  (see \cite{Grave_Camata_Coutinho_2020}). For time integration, we also use the BDF2 scheme.

\section{Navier-Stokes Equations}\label{NSE}

The Navier-Stokes equations govern the fluid flow, which leads to the following nonlinear mathematical problem to be solved: let us consider a space-time domain in which the flow takes place along the interval $[0,t_f]$ given by $\Omega \subset R^{nsd}$, where $nsd$ is the number of space dimensions.  Let $\Gamma$ denote the boundary of $\Omega$.  Find the pressure $p$ and the velocity $\mathbf{u}$ satisfying the following equations:

\begin{equation}\label{momentumadfinal}
\rho(\phi)\frac{\partial \mathbf{u}}{\partial t} +\rho(\phi)\mathbf{u} \cdotp \nabla \mathbf{u} + \nabla p - \nabla \cdotp (\mu(\phi)\nabla\mathbf{u}) - \rho(\phi)\mathbf{g} - \mathbf{F_{st}}(\phi)= 0 \textrm{ in }\Omega \times [0,t_f]
\end{equation} 

\begin{equation}\label{massad}
\nabla \cdotp \mathbf{u} = 0 \textrm{ in }\Omega \times [0,t_f].
\end{equation}

\noindent where $\rho$ is the density, $\mu$ is the dynamic viscosity and $\mathbf{F_{st}}(\phi )$ is the surface tension force, given by the Continuum Surface Model (CSF) \cite{brackbill1992continuum},

\begin{equation}\label{sf1}
%\begin{split}
\mathbf{F_{st}}(\phi)= \sigma \kappa \mathbf{n}_\phi \delta_S %\\
%&=\sigma \nabla \cdotp ((\mathbf{I} - \mathbf{n}_\phi  %\otimes \mathbf{n}_\phi) \delta_S) \\
%&= \sigma \frac{3\sqrt{2}}{4}\epsilon \nabla \cdotp (|\nabla \phi|^2 \mathbf{I} -\nabla \phi \otimes \nabla \phi)
%\end{split}
\end{equation} 

Equations \eqref{momentumadfinal} and \eqref{massad} are supplemented by proper boundary and initial conditions. For both methods is possible to evaluate the curvature and the normal vector to the interface with the phase functions: $\kappa = \nabla \cdotp \left(\frac{\nabla \phi}{||\nabla \phi||}\right)$ and $\mathbf{n}_\phi = \frac{\nabla \phi}{||\nabla \phi||}$. $\sigma$ is the surface tension coefficient and $\delta_S$ is a Dirac delta function where the surface tension is distributed.

In the Allen-Cahn phase-field method, to represent the heterogeneous flow material properties, such as the density $\rho$ and the dynamic viscosity $\mu$, we introduce to the following mixing laws,

\begin{equation}\label{rho}
\rho(\phi) = \frac{1+\phi}{2} \rho_1 +  \frac{1-\phi}{2} \rho_2
\end{equation} 

\begin{equation}\label{mu}
\mu(\phi) = \frac{1+\phi}{2} \mu_1 +  \frac{1-\phi}{2} \mu_2
\end{equation}

We consider the Dirac delta $\delta_S$ as,
 $\delta_S = \alpha_S||\nabla \phi||$ and $\alpha_S = \frac{1}{2}$, that leads to:

\begin{equation}\label{sf2}
\begin{split}
\mathbf{F_{st}}(\phi)&= \frac{1}{2} \sigma \nabla \cdotp \left(\mathbf{n}_\phi\right) \nabla \phi  
\end{split}
\end{equation} 

$\alpha_S$ is a constant derived by the property of the Dirac delta function. We find $\alpha_S$ that satisfies:

\begin{equation}
    \alpha_S\int_{-\infty}^{+\infty}  ||\nabla \phi||  dR = 1
\end{equation}

\noindent which leads to $\alpha_S = \frac{1}{2}$. $R$ is the coordinate normal to the interface. Other Dirac delta functions for phase-field models can be found in \cite{lee2012regularized}. We use the order parameter function to define the viscosity, density and surface tension. Therefore, we need a sharp transition between phases. We define $\epsilon$ between $h_e$ and $2h_e$, where $h_e$ is the minimal mesh size.

The convected level-set method defines the heterogeneous flow material properties depending on the order parameter function,

\begin{equation}
\rho = \rho_1\mathcal{H}_{scaled}(\alpha)+\rho_2(1-\mathcal{H}_{scaled}(\alpha))
\end{equation}
\begin{equation}
\mu = \mu_1\mathcal{H}_{scaled}(\alpha)+\mu_2(1-\mathcal{H}_{scaled}(\alpha))
\end{equation}

\noindent where $\mathcal{H}_{scaled}$ is a non-symmetrical, smoothed Heaviside function \cite{yokoi2014density}. Note that we need a coordinate change of the order parameter to evaluate the viscosity, density, and surface tension that follow the non-symmetrical Heaviside function. For a good coordinate change, we define $E$ between $2h_e$ and $5h_e$. The difference between the scaled ($\mathcal{H}_{scaled}$ and $\delta_{scaled}$) and the regularized ($\mathcal{H}_{\epsilon}$ and $\delta_{\epsilon}$) Heaviside and Dirac functions is that the scaled functions shift the distribution of the regularized Dirac function to the higher density region in the transition of the two phases and improves the stability of the CSF model \cite{brackbill1992continuum, bussmann2000modeling, yokoi2014density}.

\begin{equation}
\begin{split}
\mathcal{H}_{scaled}(\alpha)=  \left\{\begin{array}{ccc}
0 \textnormal{ for }  \alpha < -\epsilon_{LS} \\
\frac{1}{2}\left(\frac{1}{2}+\frac{\alpha}{\epsilon_{LS}}+ \frac{\alpha^2}{2\epsilon_{LS}^2}- \frac{1}{4\pi^2}\left(\cos \left(\frac{2\pi\alpha}{\epsilon_{LS}}\right)-1\right) + \frac{\epsilon+\alpha}{\epsilon_{LS}\pi}\sin \left(\frac{\pi\alpha}{\epsilon_{LS}}\right)\right)    \textnormal{ for }  |\alpha| \leq \epsilon_{LS}  \\
1  \textnormal{ for }  \alpha > \epsilon_{LS} \end{array}\right.
\end{split}
\label{scaled_heaviside}
\end{equation}

The Dirac function $\delta(\phi)$ satisfies $\delta(\phi)=\frac{\partial \mathcal{H}}{\partial \phi}$. Thus, we have,
 
 \begin{equation}
\begin{split}
\delta_{scaled}(\alpha)=  \left\{\begin{array}{ccc}
0 \textnormal{ for }  \alpha > \epsilon_{LS} \\
\frac{1}{2}\left(\frac{1}{\epsilon_{LS}}+ \frac{\alpha}{\epsilon_{LS}^2} + \frac{1}{2\pi\epsilon_{LS}}\sin \left(\frac{2\pi\alpha}{\epsilon_{LS}}\right) +\frac{1}{\epsilon_{LS}\pi}\sin \left(\frac{\pi\alpha}{\epsilon_{LS}}\right)+ \frac{\epsilon+\alpha}{\epsilon_{LS}^2}\cos \left(\frac{\pi\alpha}{\epsilon_{LS}}\right)\right)    \textnormal{ for }  |\alpha| \leq \epsilon_{LS} \\
0  \textnormal{ for }  \alpha < \epsilon_{LS}. \end{array}\right.
\end{split}
\label{scaled_dirac}
\end{equation}

We use a finite element Residual-Based Variational Multiscale Formulation (RBVMS) to approximate the Navier-Stokes equations and for time integration the Backward Euler method. Detailed reviews of the RBVMS formulation are in \cite{hughes,rasthofer,ahmed2017,codina2018}. See \cite{Grave_Camata_Coutinho_2020} for a complete description of the equations and methods used in this work.

\section{Implementation and Adaptive Mesh Refinement and Coarsening}\label{AMR}

All implementations are done using \texttt{libMesh}, a C++ FEM open-source software library for parallel adaptive finite element applications \cite{libmesh}. \texttt{libMesh} also interfaces with external solver packages like PETSc \cite{petsc} and Trilinos \cite{trilinos}. It provides a finite element framework that can be used for the numerical simulation of partial differential equations on serial and parallel platforms. This library is an excellent tool for programming the finite element method and can be used for one-, two-, and three-dimensional steady and transient simulations. The \texttt{libMesh} library also has an adaptive mesh refinement and coarsening strategy.

The order parameter that defines the phases has its gradients in a small region near the interface. Thus, it is not necessary a large refinement where the order parameter is constant, and because of that, we use an AMR/C strategy as follows. The AMR/C procedure uses a local error estimator considering the error of an element relative to its neighbor elements in the mesh. This error may come from any variable of the system. As it is standard in \texttt{libMesh}, Kelly$'$s error indicator is employed, which uses the H1-seminorm to estimate the error \cite{ainsworth}. Apart from the element interior residual, the flux jumps across the inter-element faces
influence the element error. The flux jump of each face is computed and added to the error contribution of the cell. For both the residual and flux jump, the values of the desired variables at each node are necessary. Being so, the error $\left\lVert e \right\rVert^2$ can be stated as,
\begin{equation}
\left\lVert e \right\rVert^2 = \sum_{i=1}^n \left\lVert e \right\rVert^2_i 
\end{equation}
\noindent where $\left\lVert e \right\rVert^2_i$ is the error of each variable. In this study, we use the order parameter function as the variable for the error estimator.

After computing the error values the elements are ''flagged” for refining and coarsening regarding their relative error. This is done by a statistical element flagging strategy. It is assumed
that the element error $\left\lVert e \right\rVert$ is distributed approximately in a normal probability function. Here, the statistical mean $\mu_s$ and standard deviation $\sigma_s$ of all errors are calculated. Whether an element
is flagged is depending on a refining ($r_f$) and a coarsening ($c_f$) fraction. For all errors $\left\lVert e \right\rVert < \mu_s - \sigma_s c_f$ the elements are flagged for coarsening and for all $\left\lVert e \right\rVert > \mu_s + \sigma_s r_f$ the elements
are marked for refinement. The refinement is performed by local isotropic subdivision (h-refinement) with hanging nodes. Here, the refinement level is limited by a maximum $h$-level
($h_{max}$) and the coarsening is done by h-restitution of sub-elements \cite{peterson}, \cite{rossa}, \cite{kelly}. 

Our algorithm starts initializing the order parameter and other variables of each interface capturing method. Then, it calls a routine to calculate the normals and curvatures. Afterward, it calculates the Navier-Stokes equation using the previously defined fluid properties. The PETSc library solves the linear system of equations coming from the linearization of the Navier-Stokes equations invoked by \texttt{libMesh}, applying the GMRES with Block-Jacobi preconditioner together with ILU(0) within each block. Then, the interface movement is calculated with the convected level-set method or the Allen-Cahn phase-field method using the velocities coming from the fluid flow model. The resulting linear system of the interface capturing equations is solved in the same way as the fluid flow model. For the convected level-set, the global mass conservation is then enforced. With the flow and order parameter updated, the adaptive mesh refinement algorithm is applied, refining or coarsening the mesh depending on the flux jump of the order parameter. Finally, the flow and interface capturing equations are recalculated with the updated mesh before going for the next time step.

\section{Numerical results}\label{Results}

This section presents the application of both interface-capturing methods in relevant bubble numerical simulations. The implementation of the convected level-set was already verified in \cite{Grave_Camata_Coutinho_2020}. Thus, we first focus on validating the implementation of the Allen-Cahn equations. Then we present the comparison of the simulations of a rising bubble and droplet impact using these methods considering AMR/C.

The parameters of each problem are the same for both interface-capturing methods, convected level-set, and Allen-Cahn phase-field. The only difference is how we model the interface's motion.

\subsection{Verification of the Allen–Cahn implementation}

We first verify the Allen–Cahn solver using the volume-conserved motion by curvature in two dimensions. A square computational domain [0, 1]$\times$[0, 1] is considered.

\begin{figure}[ht!]
    \centering
    \begin{minipage}{.45\textwidth}
    \includegraphics[width=0.8\linewidth]{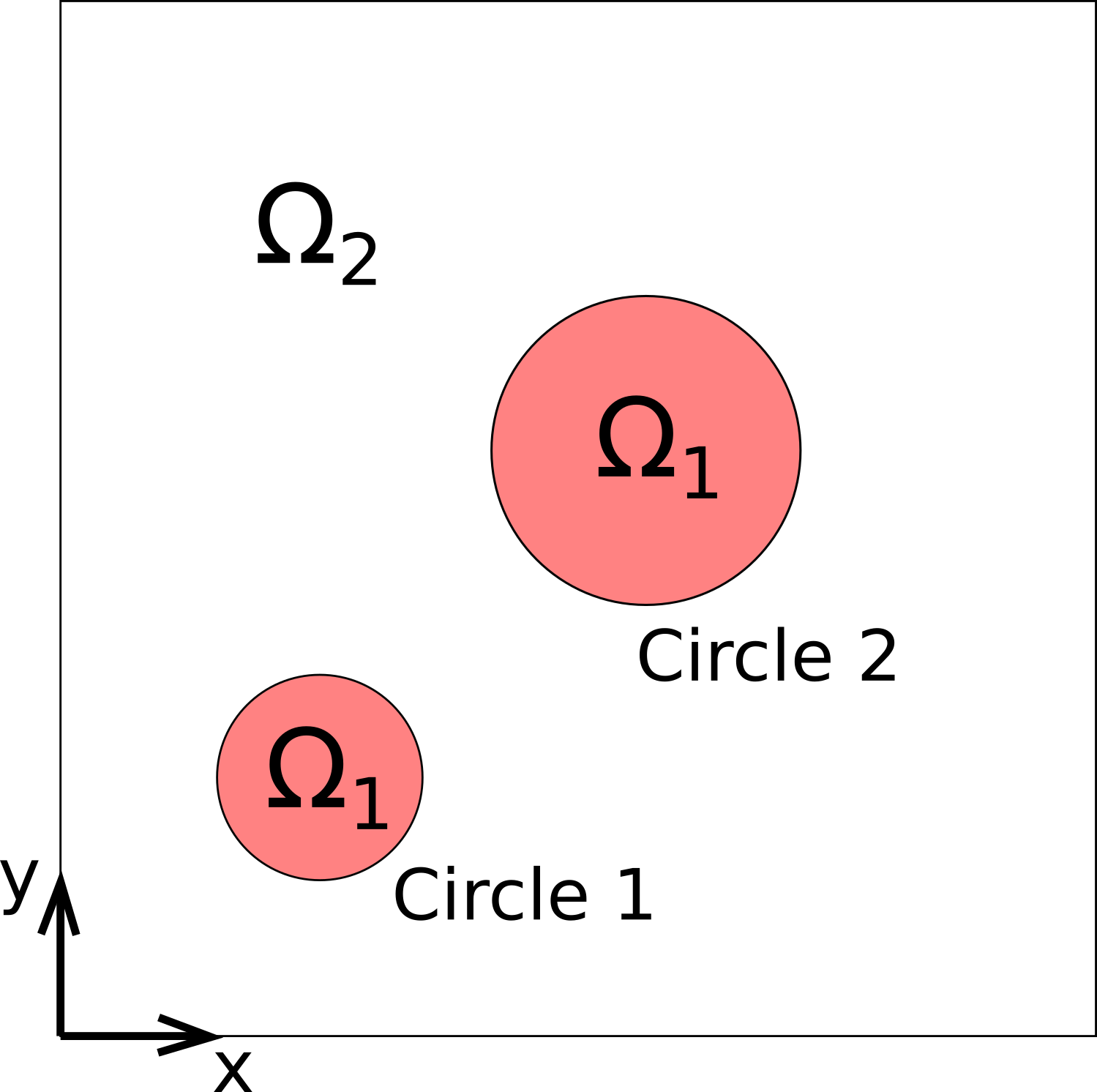}
    \caption{Allen-Cahn validation: Schematic diagram.}
    \label{fig:verification}
\end{minipage}
\begin{minipage}{.45\textwidth}
    \centering
    \includegraphics[width=\linewidth]{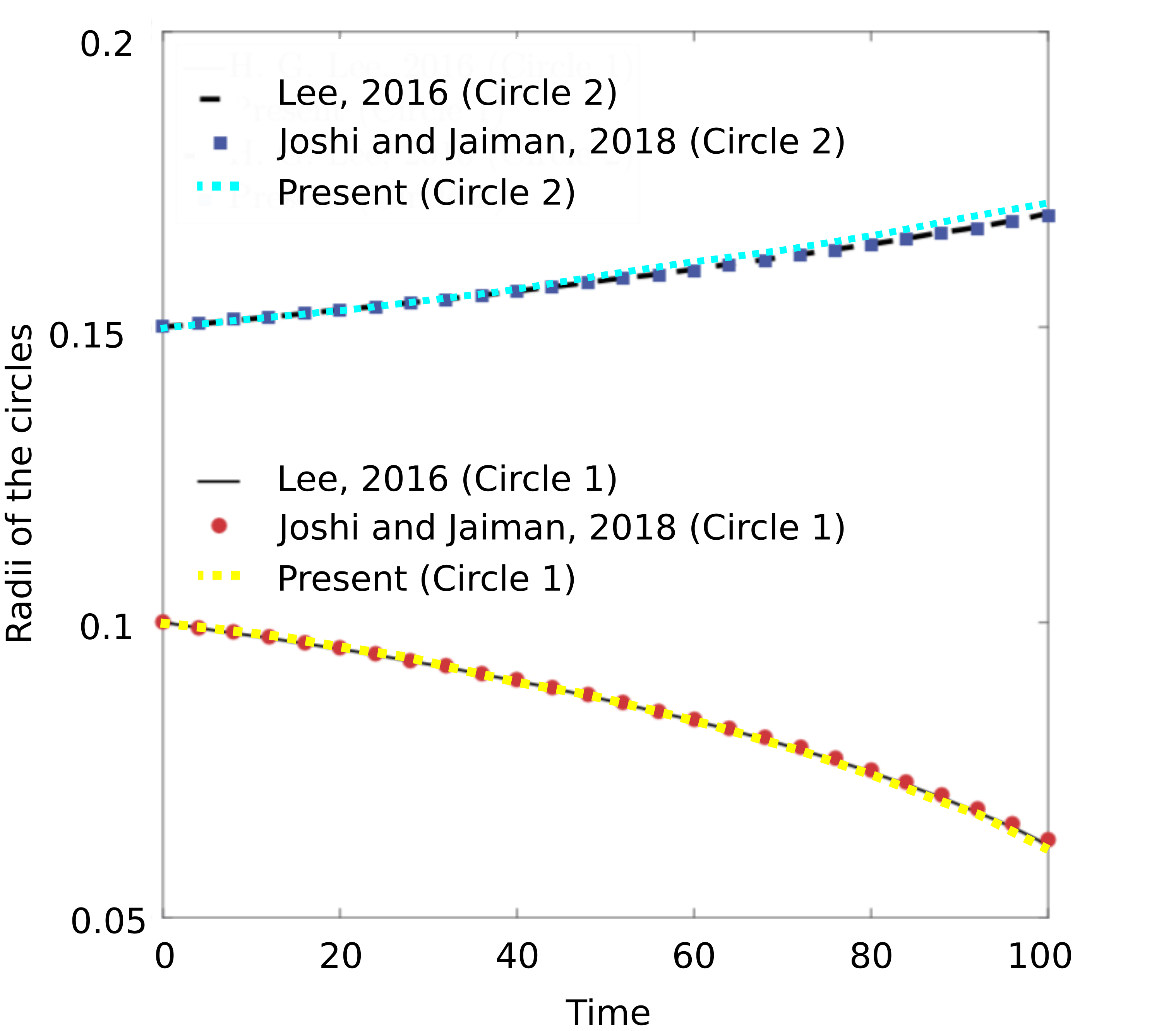}
    \caption{Allen-Cahn validation: Validation of the evolution of the
radii of the two circles with the literature \cite{joshi2018positivity,lee2016high} }
    \label{fig:results_verification}
    \end{minipage}
\end{figure}

\begin{figure}[ht!]
    \centering
    \includegraphics[width=\linewidth]{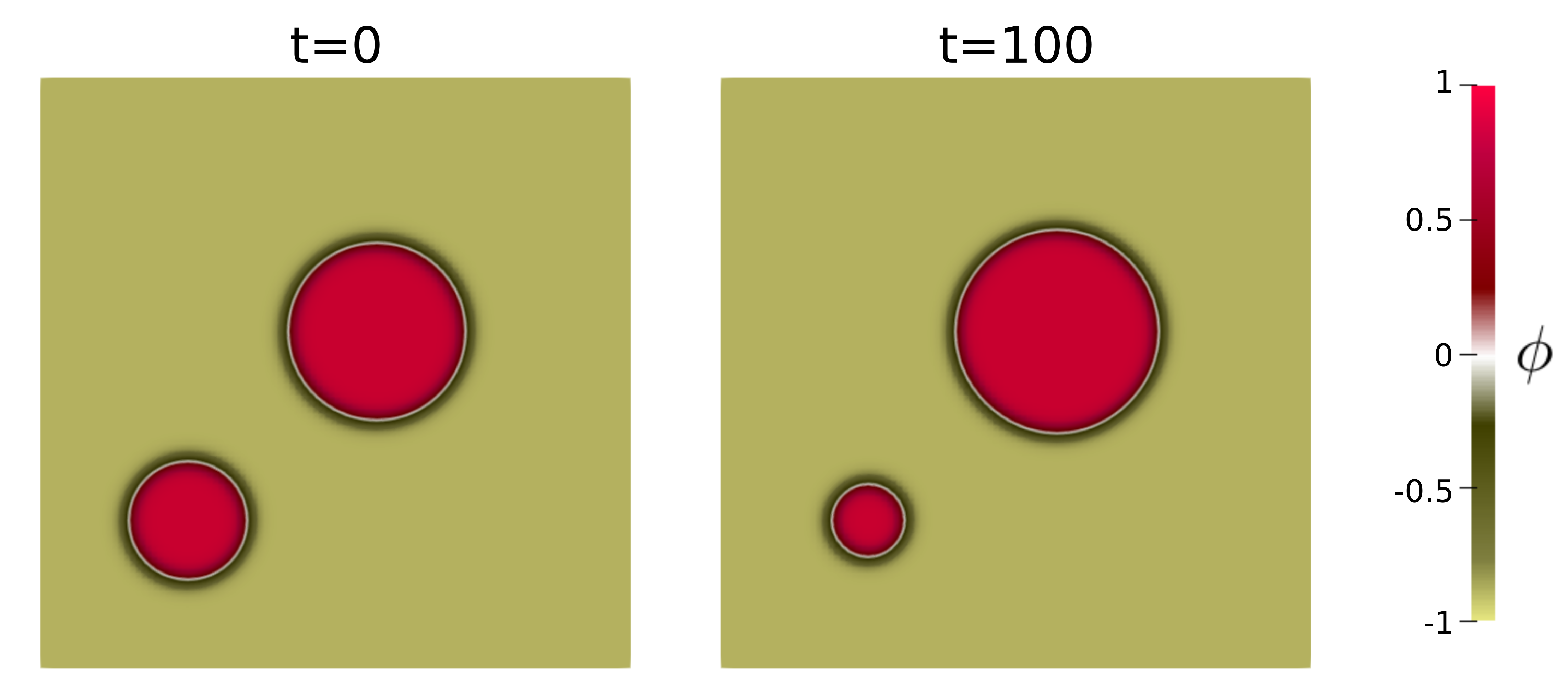}
    \caption{Contour plots of the order parameter $\phi$ at t = 0, and t = 100}
    \label{fig:comparison_t}
\end{figure}

The initial condition is given by:
\begin{equation}\label{initialcond}
\phi = 1.0 + \tanh \left(\frac{R_1 - \sqrt{(x-0.25)^2 + (y-0.25)^2}}{\sqrt{2}\epsilon}\right) +\tanh \left(\frac{R_2 - \sqrt{(x-0.57)^2 + (y-0.55)^2}}{\sqrt{2}\epsilon}\right)
\end{equation} 

\noindent where $R_1$ = 0.1 and $R_2$ = 0.15 are the radii of the two circles centered at (0.25, 0.25) and (0.57, 0.57), respectively. A schematic diagram of the problem is shown in Fig. \ref{fig:verification}. The variation of the change of the radii of the two circles is tracked and validated with the results obtained in \cite{mao2021variational} for $\epsilon = 0.01$ (Fig. \ref{fig:results_verification}).
The time-step size in the present simulation is 0.1 with the final time t = 100. The problem set-up with the evolution of the radii of the two circles is shown in Fig. \ref{fig:comparison_t}. The results are in very close agreement with the reference.

\subsection{Laplace-Young Law}

To verify the coupling between the Allen–Cahn and the incompressible Navier–Stokes equations at a high density ratio, we consider a simple test problem of Laplace–Young law. The law states that the pressure difference ($\Delta p$) across the interface
of a static bubble in a two-phase fluid system is equivalent to the ratio of the surface tension ($\sigma$) and the radius of curvature
($R$) of the bubble,

\begin{equation}
    \Delta p = \frac{\sigma}{R}
\end{equation}

The density and dynamic viscosity of the two fluids are taken as $\rho_1$ = 1000, $\rho_2$ = 1, $\mu_1$ = 10 and $\mu_2$ = 0.1. In the numerical tests, we consider a domain size  = [0.4] $\times$ [0.4] with uniform structured mesh of grid size 1/200 with different
radii of the bubble (0.2, 0.4 and 0.6 units) and three surface tensions (0.5, 0.25 and 0.05 units).

The initial condition is given by:

\begin{equation}\label{initialcond2}
\phi = -\tanh \left(\frac{R - \sqrt{(x-2)^2 + (y-2)^2}}{\sqrt{2}\epsilon}\right)
\end{equation} 

\begin{figure}[!ht]
  \centering
  \subfloat[Laplace-Young Law: Schematic diagram.]{\includegraphics[width=0.5\textwidth]{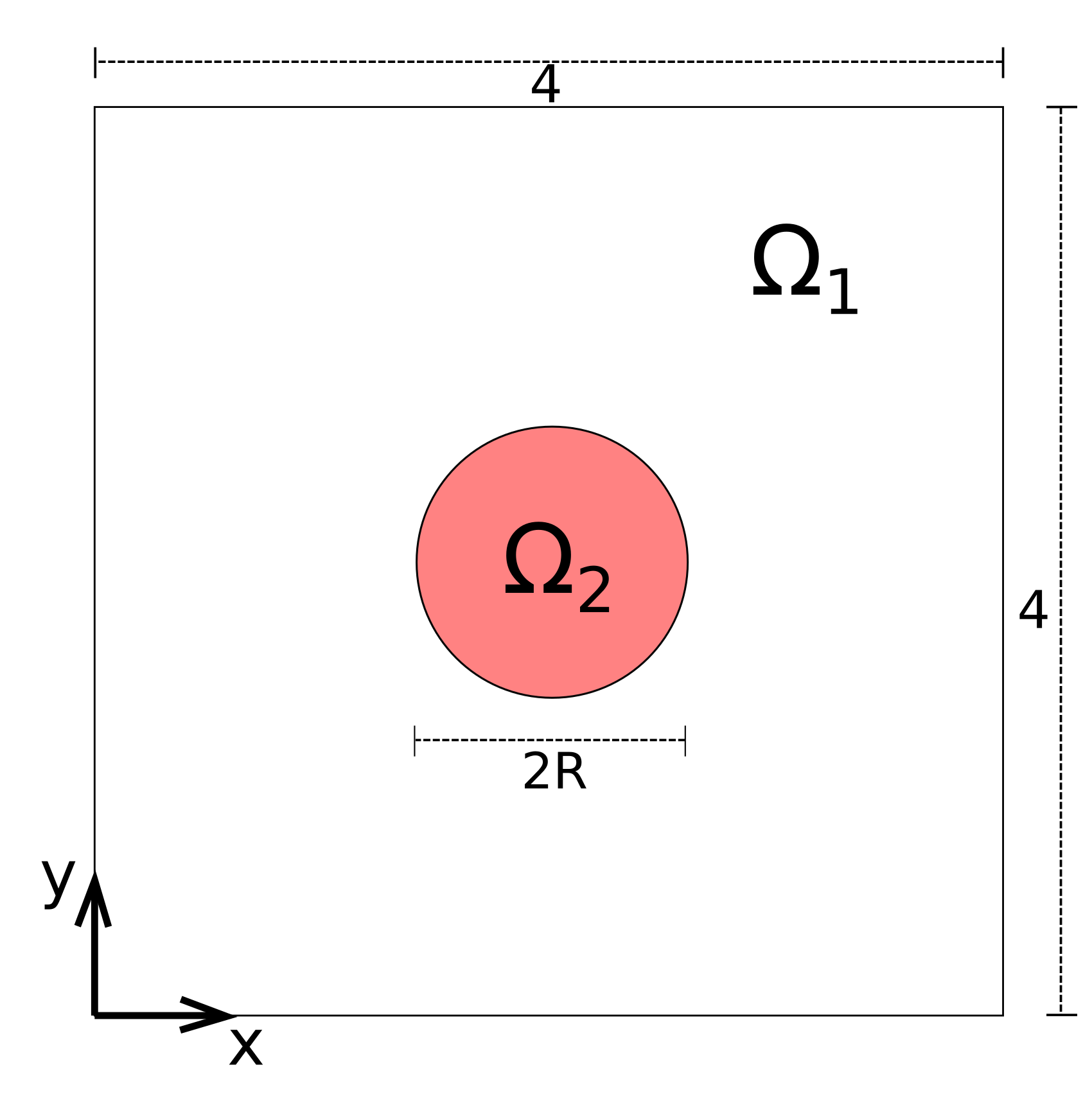}\label{schematic_pressure}}
  \hfill
  \subfloat[R=0.2.]{\includegraphics[width=0.5\textwidth]{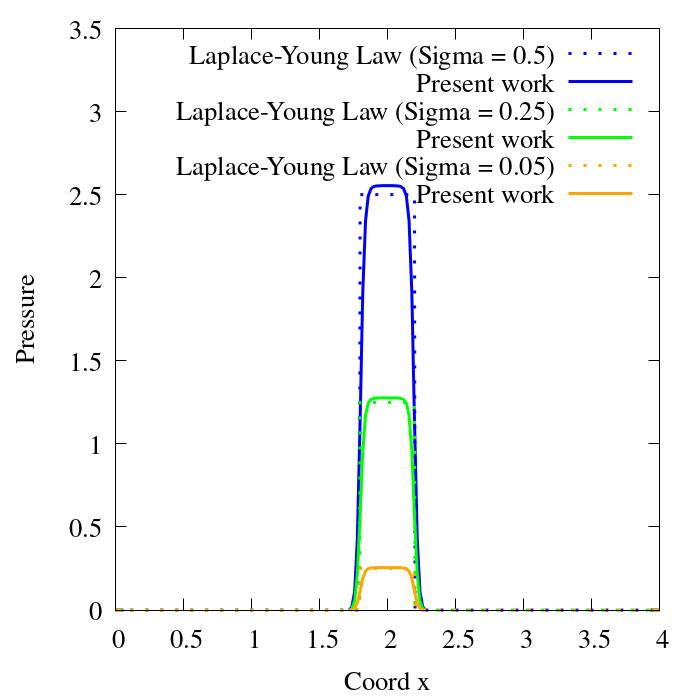}\label{r02}}
    \hfill
  \subfloat[R=0.4.]{\includegraphics[width=0.5\textwidth]{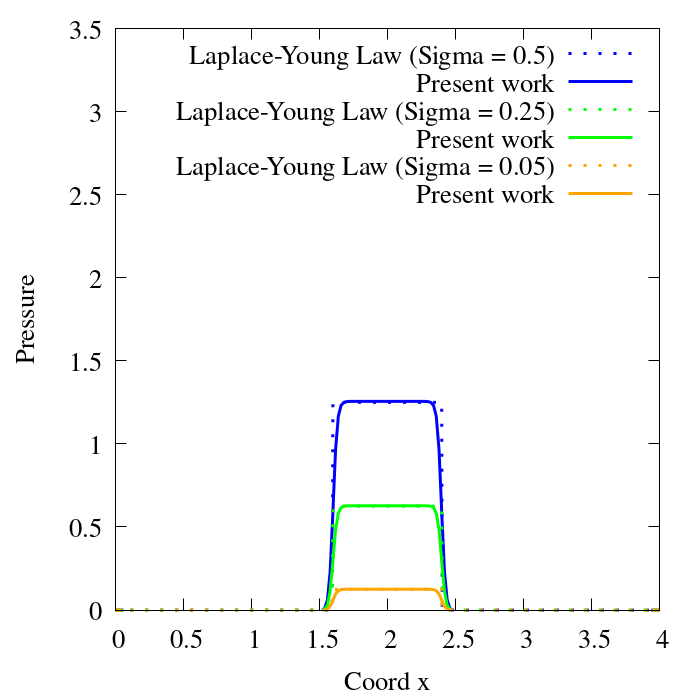}\label{r04}}
    \hfill
  \subfloat[R=0.6.]{\includegraphics[width=0.5\textwidth]{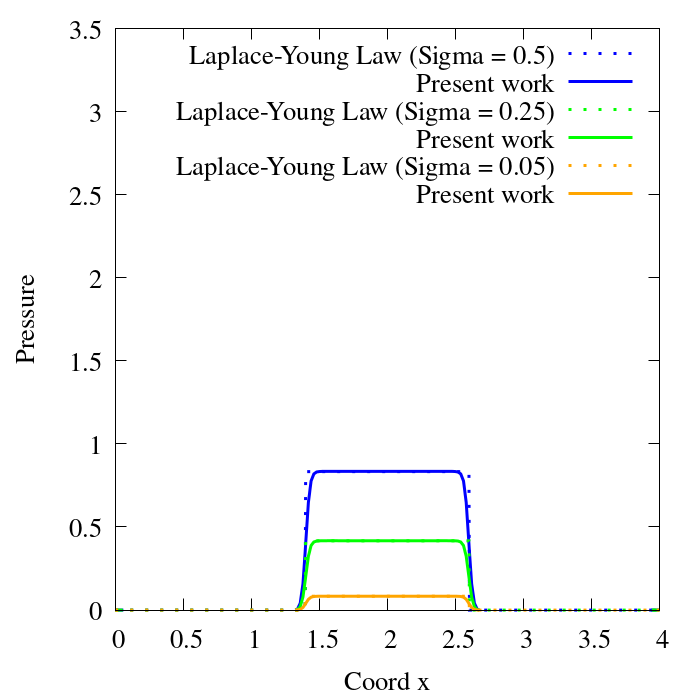}\label{r06}}
  \caption{Laplace-Young Law.}
  \label{laplace_young}
\end{figure}

The interface thickness parameter is $\epsilon = 0.02$ and $\eta = 0.05$ %($\eta = 0.002$ caused oscillations).
The time-step size is taken as $\Delta t$ = 0.1 s, and the pressure difference is measured after 500 time-steps. Since the first iteration, the pressure difference shows good results and keeps the same values until the end of the simulation. The schematic of the computational domain and the representative results are shown in Fig. \ref{laplace_young}. The pressure difference shows good agreement with the Laplace–Young law. Table \ref{tab:laplace_error} shows the relative error between the pressure difference of each simulation, which decreases as the radii increases. For $R=0.2$ the error is about 2\%, $R=0.4$ about 0.5\% and $R=0.6$ about 0.2\%. These low relative errors suggest that the coupling between the Allen–Cahn and the Navier–Stokes equations is verified for a high density ratio problem.

\begin{table}[ht!]
  \begin{center}
  \caption{Laplace-Young Law: Relative error.}
   \label{tab:laplace_error}
    \begin{tabular}{c c c c} % <-- Alignments: 1st column left, 2nd middle and 3rd right, with vertical lines in between
      \hline
Radii & $\sigma=0.5$ & $\sigma=0.25$ & $\sigma=0.05$ 
\\
0.2 & 0.02174 & 0.02177 & 0.02179
\\
0.4 & 0.00490 & 0.00491 & 0.00491
\\
0.6 & 0.00196 & 0.00194 & 0.00192
\\
     \hline
    \end{tabular}
  \end{center}
  \end{table}

One concern about this test case is the spurious currents that the surface tension representation may generate. A sharp surface tension force can give rise to large spurious currents and oscillations in the pressure. On the other hand, a regularized surface tension may provide exact force balance at the interface. However, the numerical curvature calculation introduces errors that also cause spurious currents \cite{zahedi2012spurious}. The magnitude of the spurious velocities depends mainly on the density ratio between the phases and how we define the Dirac function used to model the surface tension and the coupling between fluids. Since we are using a high density ratio in this test case, the maximum value of the velocities is the order of $10^{-5}$. The spurious velocities diminish by augmenting $\epsilon$ or diminishing the density ratio. Fortunately, these spurious velocities do not seem to appear or be significant in the remaining test cases, where gravity is present. Figure \ref{fig:spurious} shows the magnitude of the spurious velocities for $\sigma=0.5$ and $R=0.6$:

\begin{figure}[ht!]
    \centering
    \includegraphics[width=0.6\linewidth]{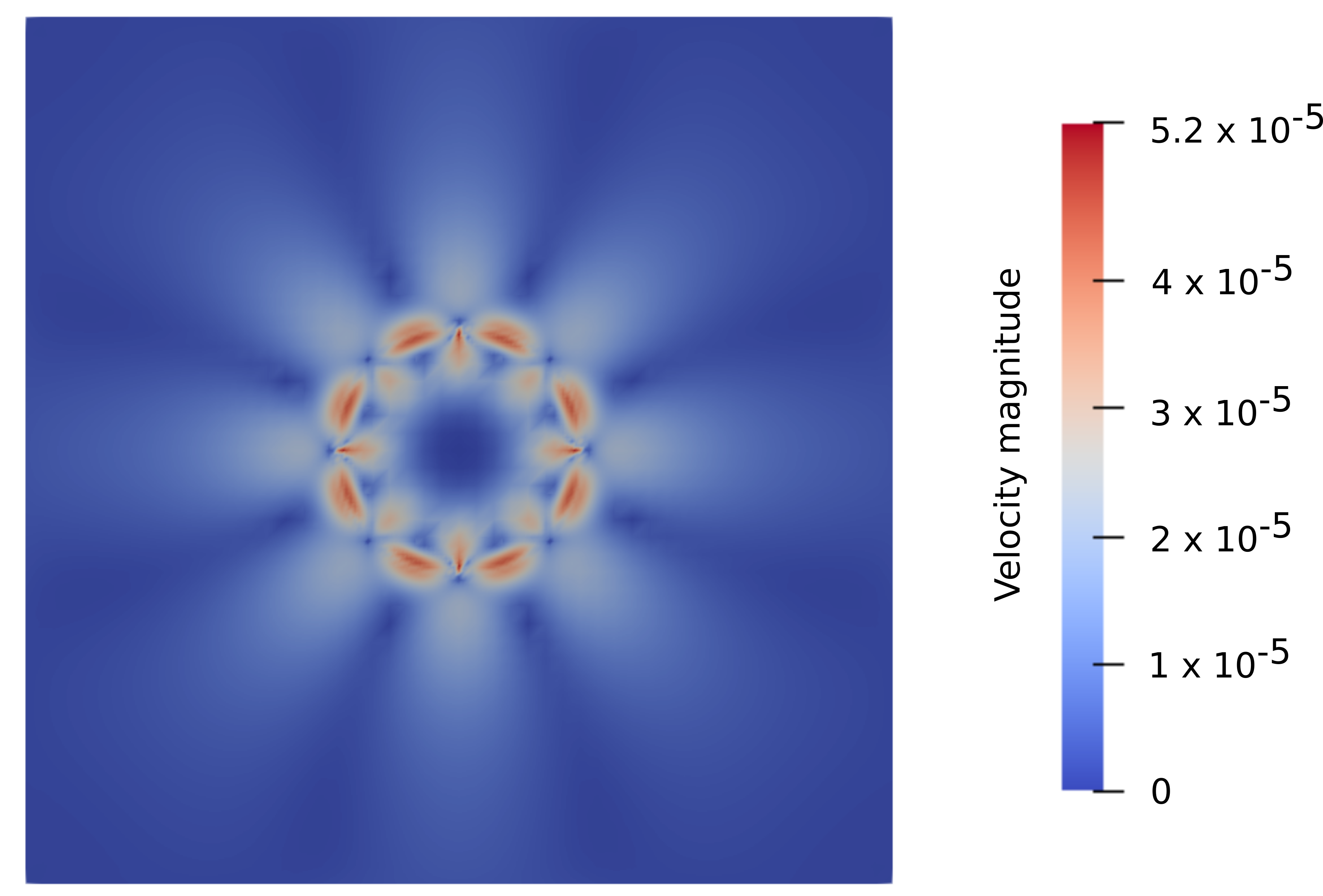}
    \caption{Laplace–Young law: Spurious velocities.}
    \label{fig:spurious}
\end{figure}

\begin{figure}[ht!]
  \centering
  \subfloat[Phase-field]{\includegraphics[width=0.5\textwidth]{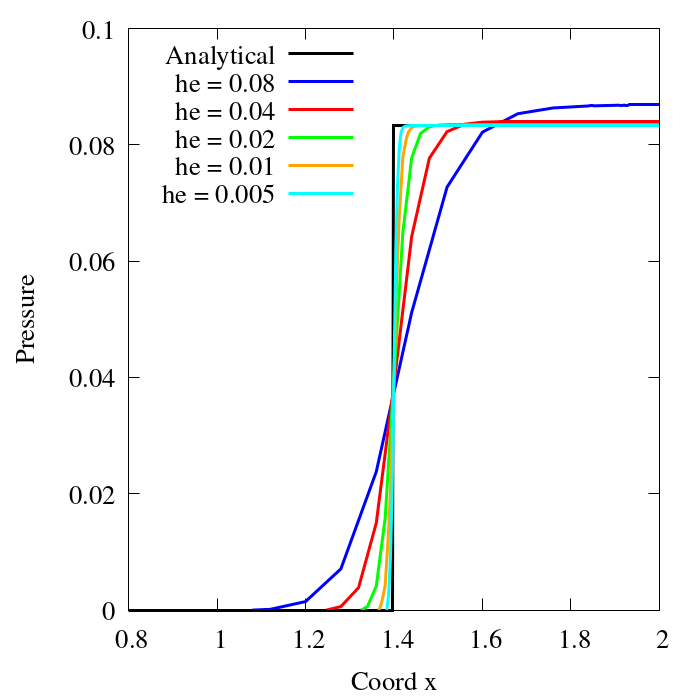}\label{fig:meshrefanalysis}}
  \hfill
  \subfloat[Convected level-set.]{\includegraphics[width=0.5\textwidth]{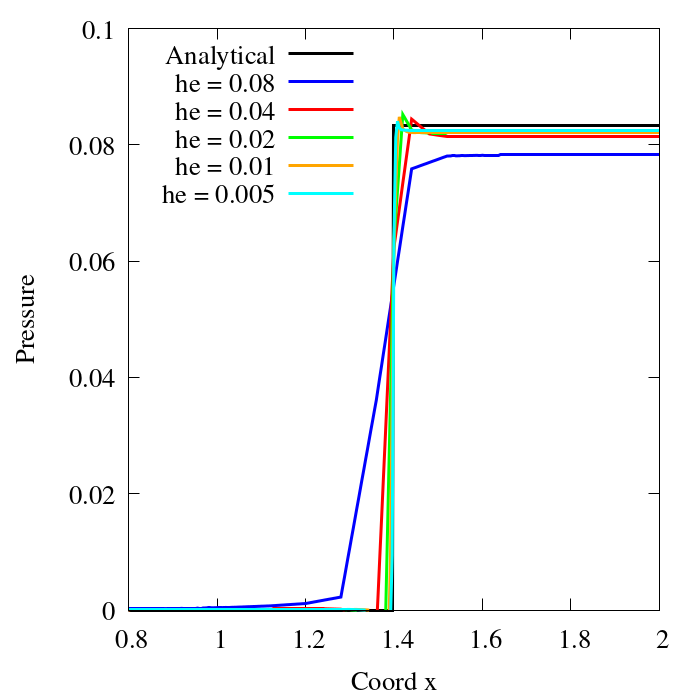}\label{fig:meshrefanalysisls}}
  \caption{Laplace–Young law: Mesh size analysis.}
\end{figure}

We also present a mesh refinement analysis for the case with $R=0.6$ and $\sigma = 0.05$ for the Allen-Cahn phase-field and convected level-set methods. We use five different mesh sizes, and we set the parameter $\epsilon = 2h_e$ for the phase-field and $E = 5h_e$ for the convected level-set, where $h_e$ is the mesh size. In Figure \ref{fig:meshrefanalysis} we see that as the mesh size decreases, we obtain a sharper change of pressure with the phase-field method, and the pressure tends to converge to the analytical value. The same happens by using the convected level-set method (Fig. \ref{fig:meshrefanalysisls}). However, the transition between pressures is sharper than the results obtained with the phase-field for all mesh sizes, even when using large elements.

%\begin{figure}[ht!]
%    \centering
%    \includegraphics[width=0.6\linewidth]{figures/r06refined.png}
%    \caption{Laplace–Young law: Mesh size analysis.}
%    \label{fig:meshrefanalysis}
%\end{figure}

\subsection{Rising bubbles}

Two well-known two-phase flow benchmarks are the 2D and 3D rising bubbles. We use these benchmarks to compare results obtained with the phase-field and level-set methods.

\subsubsection{2D rising bubble}

\begin{figure}[ht!]
\begin{center}
\includegraphics[width=0.3\textwidth]{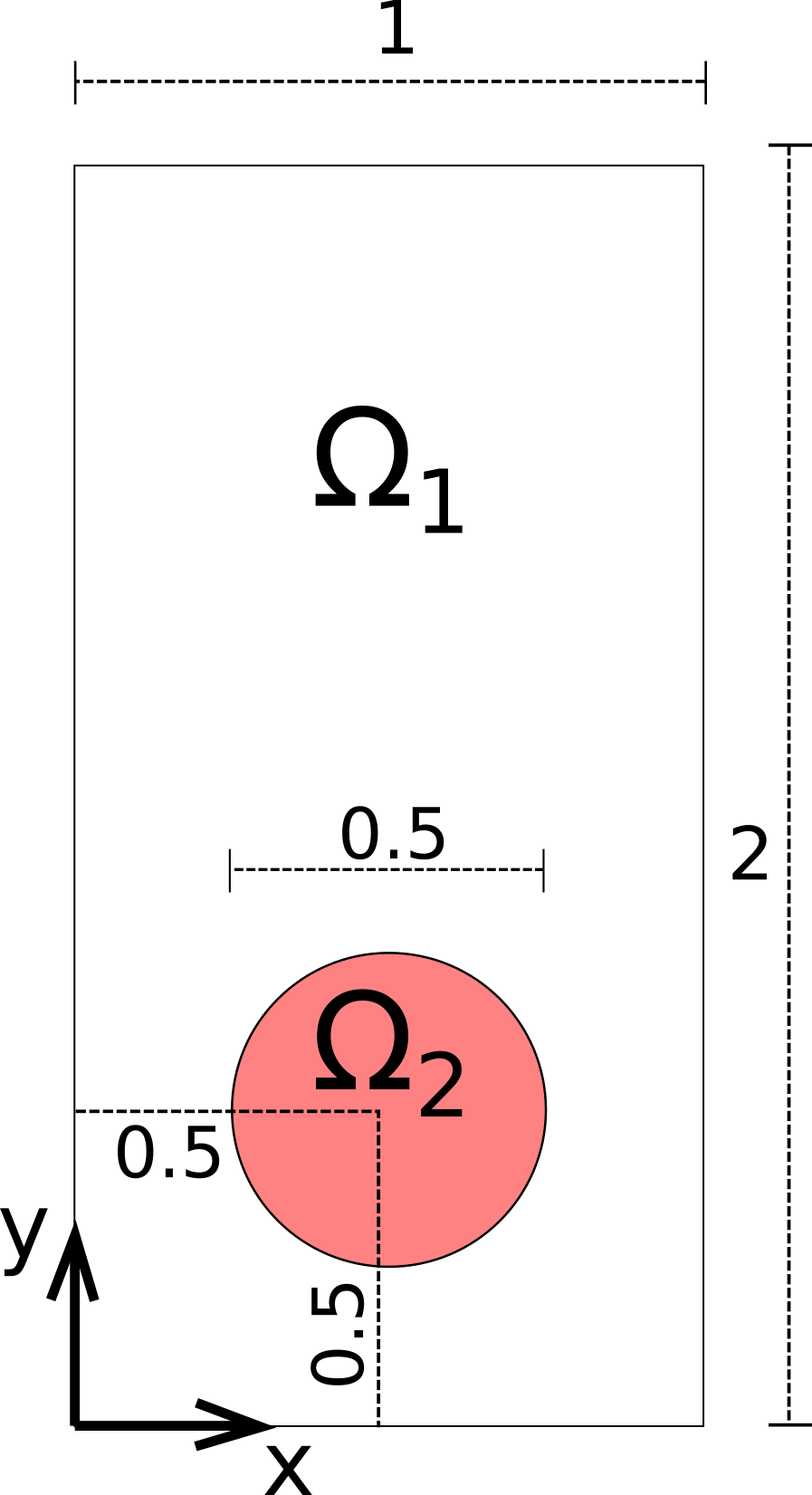}
\end{center}
\caption{2D rising bubble: Schematic diagram.}
\label{bubble_bench}
\end{figure}

We simulate two benchmarks proposed by \cite{hysing2007proposal} (case A and case B). The task of the proposed benchmarks is to track the evolution of a two-dimensional bubble rising in a liquid column, with the initial configuration described in Fig. \ref{bubble_bench}.

The initial configuration is identical for both test cases and consists of a circular bubble of radius $R = 0.25$ m centered at $[0.5, 0.5]$ m in a $[1 \times 2]$ m rectangular domain. The no-slip boundary condition is used at the top and bottom boundaries, whereas the free-slip condition is imposed on the vertical walls. Table \ref{tab:bubble} lists the parameters used for this simulation. 

\begin{table}[ht!]
  \begin{center}
  \caption{2D rising bubble: Data.}
   \label{tab:bubble}
    \begin{tabular}{c c c c} % <-- Alignments: 1st column left, 2nd middle and 3rd right, with vertical lines in between
      \hline
Computational domain & \multicolumn{2}{c}{$1\times 2$} & (m)
\\
Grid sizes & \multicolumn{2}{c}{0.05 to 0.003125} & (m)
\\
Number of time-steps & \multicolumn{2}{c}{960} & (-)
\\
Time-step & \multicolumn{2}{c}{0.005} & s
\\
Bubble radius & \multicolumn{2}{c}{0.25} & m
\\
Initial bubble position & \multicolumn{2}{c}{$(x,y) = (0.5, 0.5)$} & m
\\
Liquid density & (A) 1000 & (B) 1000 & kg/$\text{m}^3$
\\
Liquid viscosity & (A) 10 & (B) 10 & kg/(ms)
\\
Gas density & (A) 100 & (B) 1 & kg/$\text{m}^3$
\\
Gas viscosity & (A) 1 & (B) 0.1 & kg/(ms)
\\
Surface tension & (A) 24.5 & (B) 1.96 & N/m
\\
Gravity & (A) 0.98 & (B) 0.98 & m/$\text{s}^2$
\\
     \hline
    \end{tabular}
  \end{center}
  \end{table}

For the convected level-set simulation we define $\lambda = 1$ and $E = 0.015625$. For the  Allen-Cahn phase-field simulation we define $\epsilon = 0.00625$ and $\eta = 0.025$. We use an adapted mesh, initially with 20 $\times$ 40 bilinear quadrilateral elements, and after the refinement, the smallest element has a size of 0.003125 m. We refine initially the region where the bubble is located in three levels. The adaptive mesh refinement is based on the flux jump of the order parameter error, in which $h_{max} = 4$. We apply the adaptive mesh refinement every 10 time steps.

\begin{figure}[!ht]
  \centering
  \subfloat[Bubble shape at $t=3$.]{\includegraphics[width=0.5\textwidth]{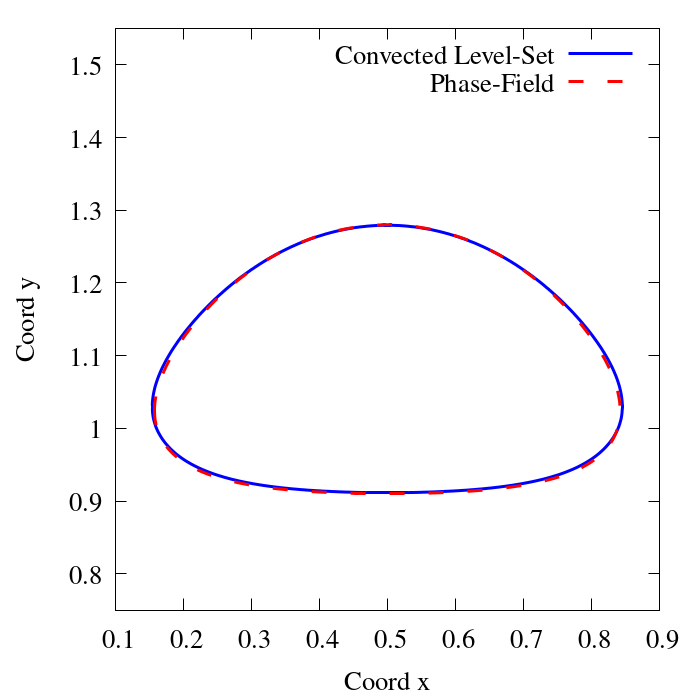}\label{benchmark1}}
  \hfill
  \subfloat[Circularity.]{\includegraphics[width=0.5\textwidth]{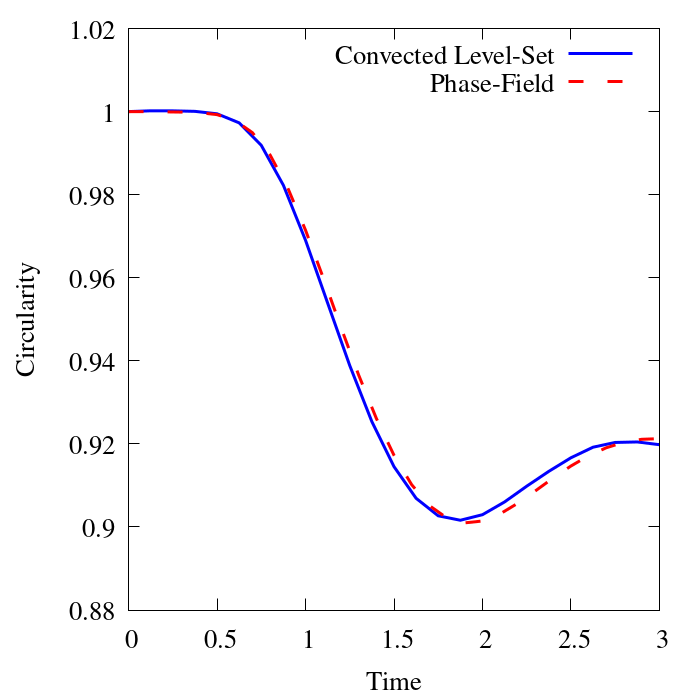}\label{circularity1}}
    \hfill
  \subfloat[Center of mass.]{\includegraphics[width=0.5\textwidth]{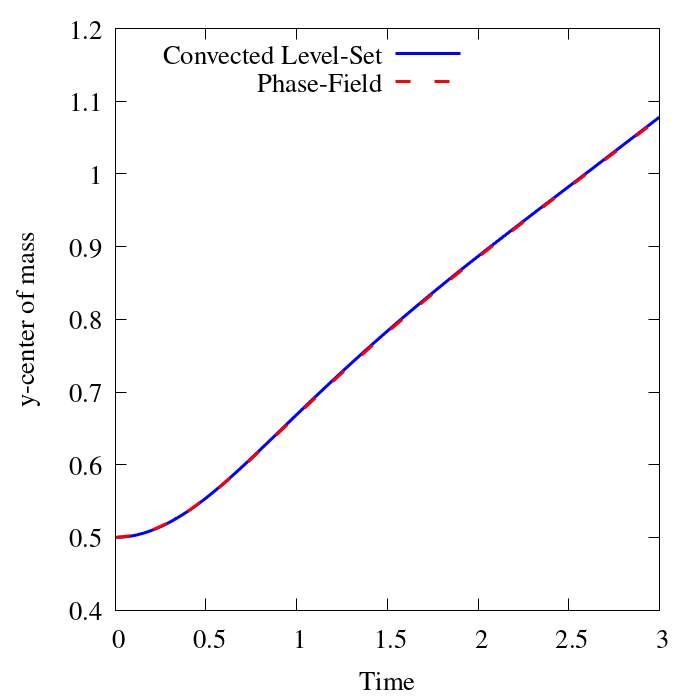}\label{mass1}}
    \hfill
  \subfloat[Rise velocity.]{\includegraphics[width=0.5\textwidth]{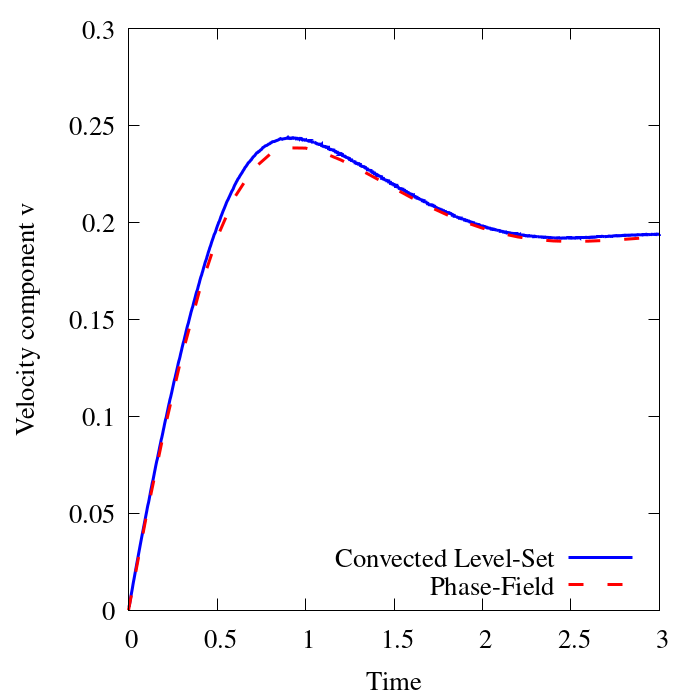}\label{risevel1}}
  \caption{2D rising bubble: Test A results.}
\end{figure}

In Fig. \ref{benchmark1}, we present the bubble shape at the final time $(t=3)$ for the test case A. The results of the convected level-set were verified against the literature in \cite{Grave_Camata_Coutinho_2020}. The prediction of the phase-field method is also in good agreement with the reference (Fig. \ref{benchmark1}). However, the final shape is not sufficient to validate the code. Therefore, we introduce the following quantities of interest, which will be used to assist in describing the temporal evolution of the bubbles quantitatively: center of mass $\mathbf{X_c}$, circularity $\slashed{c}$ and rise velocity $\mathbf{U_{mean}}$.

\begin{equation}
\mathbf{X_c} = \frac{\int_{\Omega_2}\mathbf{x} dx}{V_{\Omega_2}}.
\end{equation}

\begin{equation}
\slashed{c} = \frac{P_a}{P_b}=\frac{\textrm{perimeter of a area-equivalent circle}}{\textrm{perimeter of the bubble}} =\frac{2\pi R}{P_b}.
\end{equation}

\begin{equation}
\mathbf{U_{mean}} = \frac{\int_{\Omega_2}\mathbf{u} dx}{V_{\Omega_2}}.
\end{equation}

\noindent where $\Omega_2$ denotes the region that the bubble occupies.

In Figures \ref{circularity1}, \ref{mass1}, \ref{risevel1} we compare the circularity, center of mass position, and rise velocity for test case A. All groups have a good agreement for the quantities of interest. However, in \cite{Grave_Camata_Coutinho_2020} we may note that the curve fitting between the convected level-set and other authors was almost perfect. Here, the Allen-Cahn phase-field method returns a good prediction but does not fit perfectly with the convected level-set (which could be interpreted as a reference). Therefore, we may infer that the Allen-Cahn phase-field method needs more levels of refinement than the convected level-set to capture the same physics, since the interface thickness $\epsilon$ depends on the mesh size and the Allen-Cahn phase-field needs sharper transitions to reproduce better the effects of pressure, as we have seen in Fig. \ref{laplace_young}.

\begin{figure}[ht!]
  \centering
  \subfloat[Bubble shape at $t=3$.]{\includegraphics[width=0.5\textwidth]{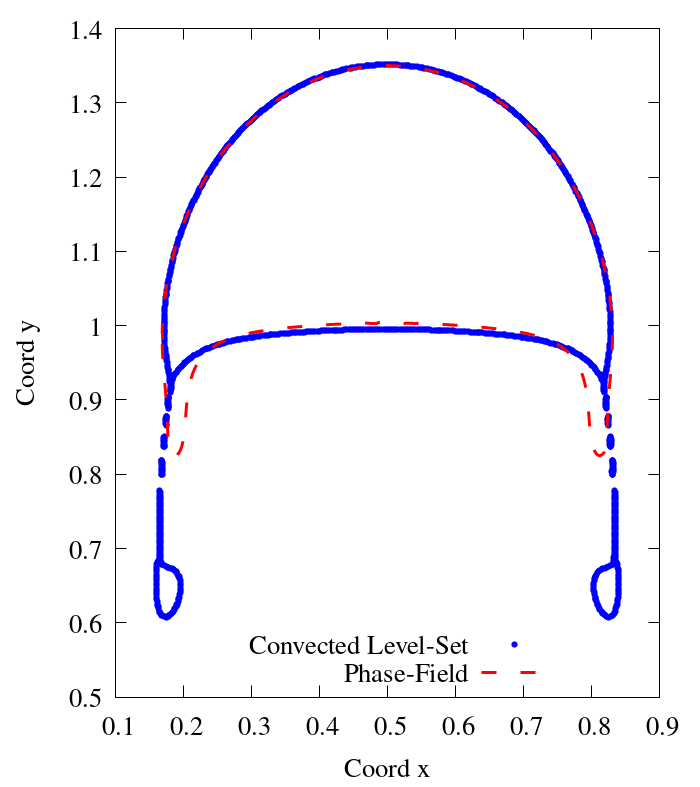}\label{benchmark2}}
  \hfill
  \subfloat[Circularity.]{\includegraphics[width=0.5\textwidth]{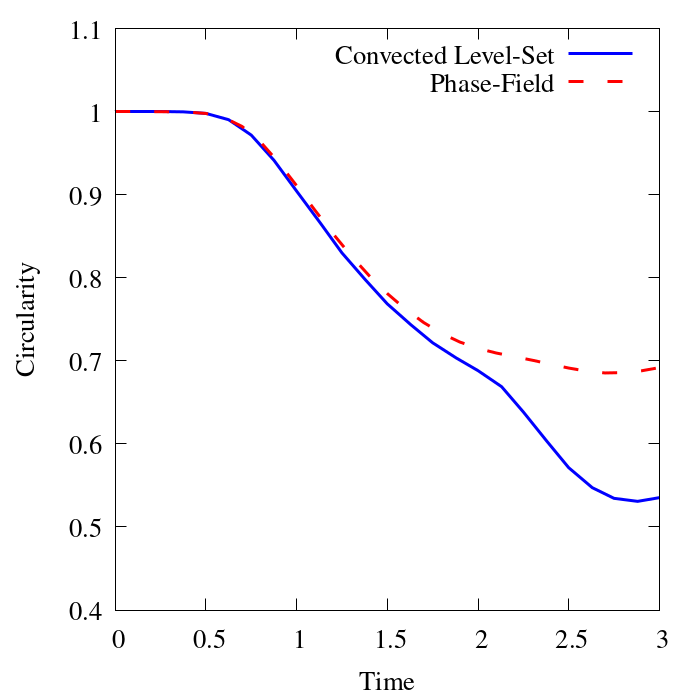}\label{circularity2}}
    \hfill
  \subfloat[Center of mass.]{\includegraphics[width=0.5\textwidth]{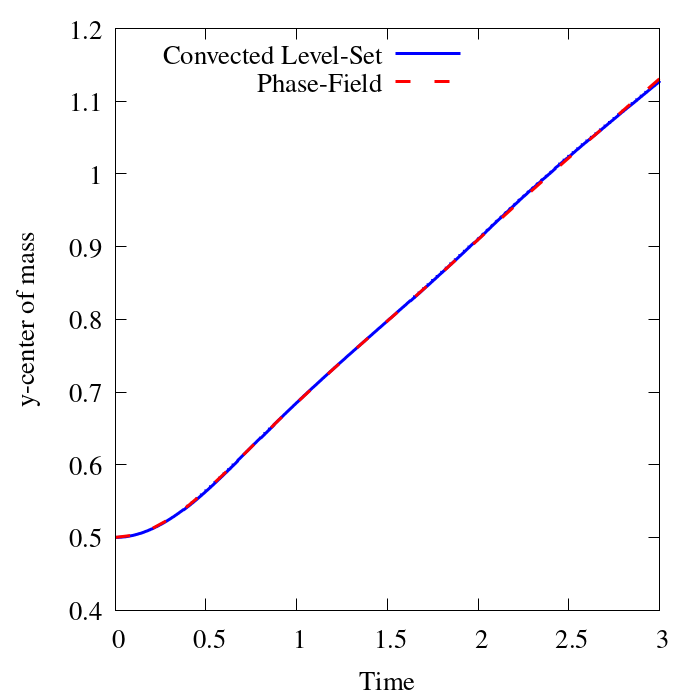}\label{mass2}}
    \hfill
  \subfloat[Rise velocity.]{\includegraphics[width=0.5\textwidth]{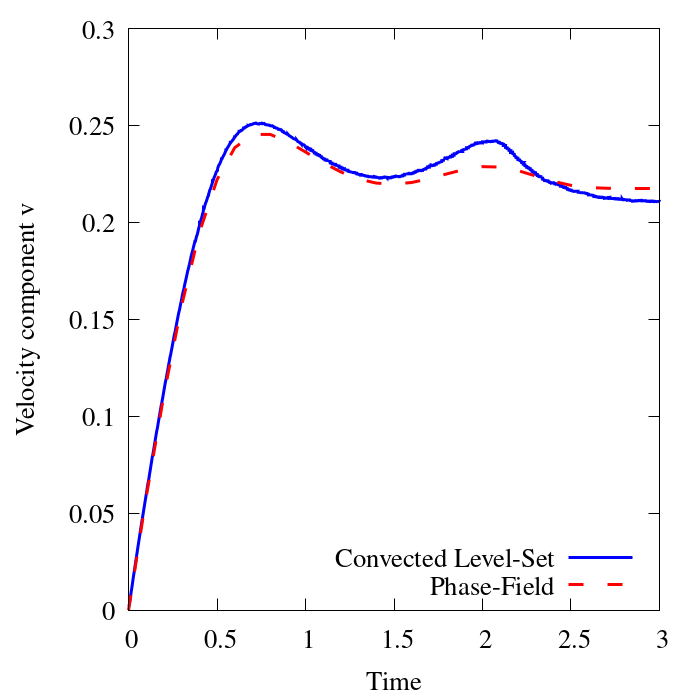}\label{risevel2}}
  \caption{2D rising bubble: Test B results.}
\end{figure}

In Fig. \ref{benchmark2}, we present the bubble shape at the final time $(t=3)$ for the test case B. Again, we compare the results between both methods. As shown in \cite{Grave_Camata_Coutinho_2020}, although both methods predict a similar shape for the main bulk of the bubble, there is no agreement concerning the thin filamentary regions. There are discrepancies when we compare all quantities of interest. The circularity (Fig. \ref{circularity2}) agrees very well until about $t=1.75$ seconds, and for later times significant differences start to appear, that is, when the thin filaments are present. The center of mass, shown in Fig. \ref{mass2}, is predicted similarly despite the shape differences, and the mean rise velocity also presents a pretty good agreement between the two methods (Fig. \ref{risevel2}).  

\begin{figure}[ht!]
  \centering
  \subfloat[Constant mobility coefficient $\gamma = 1$.]{\includegraphics[width=0.45\textwidth]{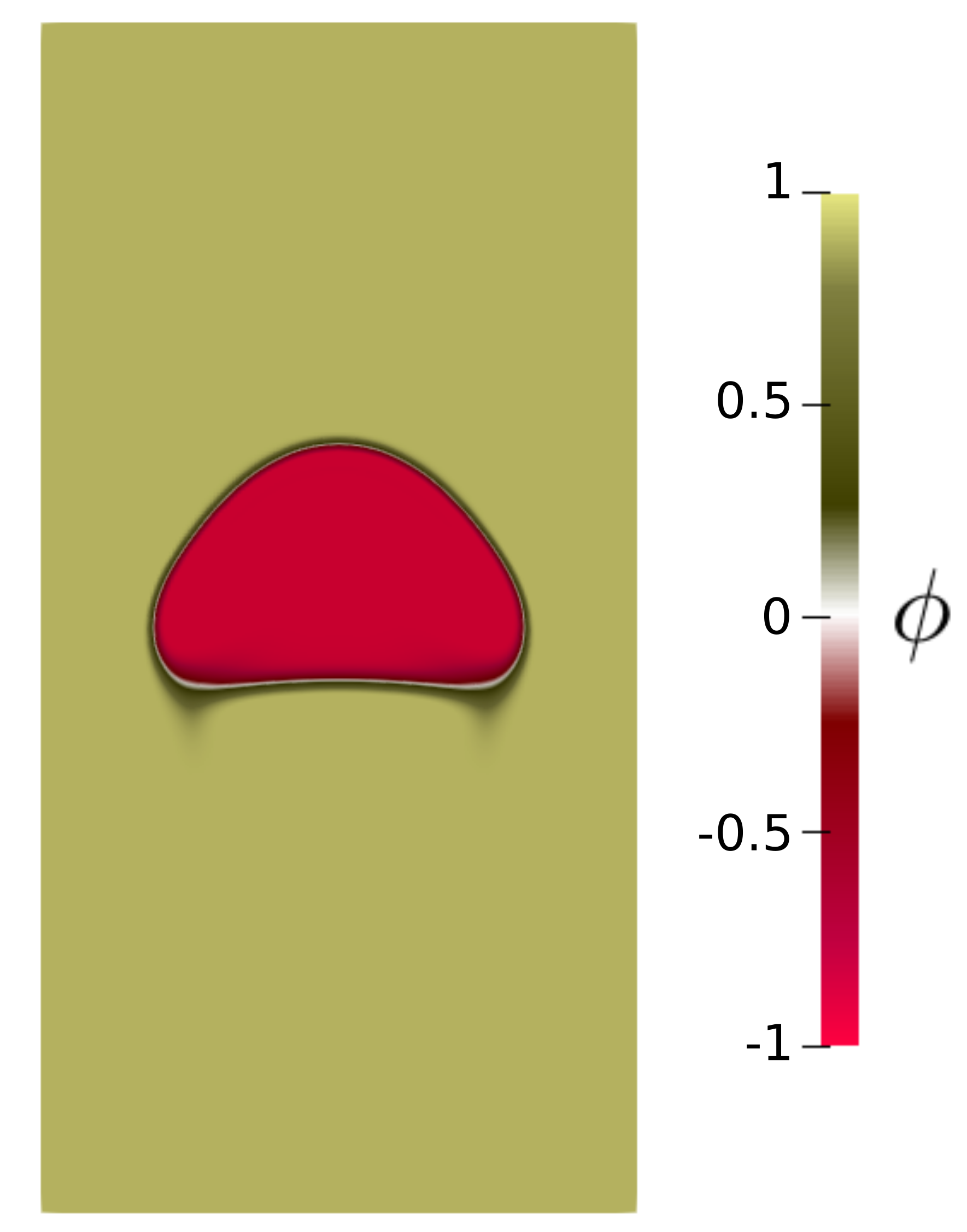}\label{mobilityfixed}}
  \hfill
  \subfloat[Time-dependent mobility model at $\eta = 0.025$.]{\includegraphics[width=0.45\textwidth]{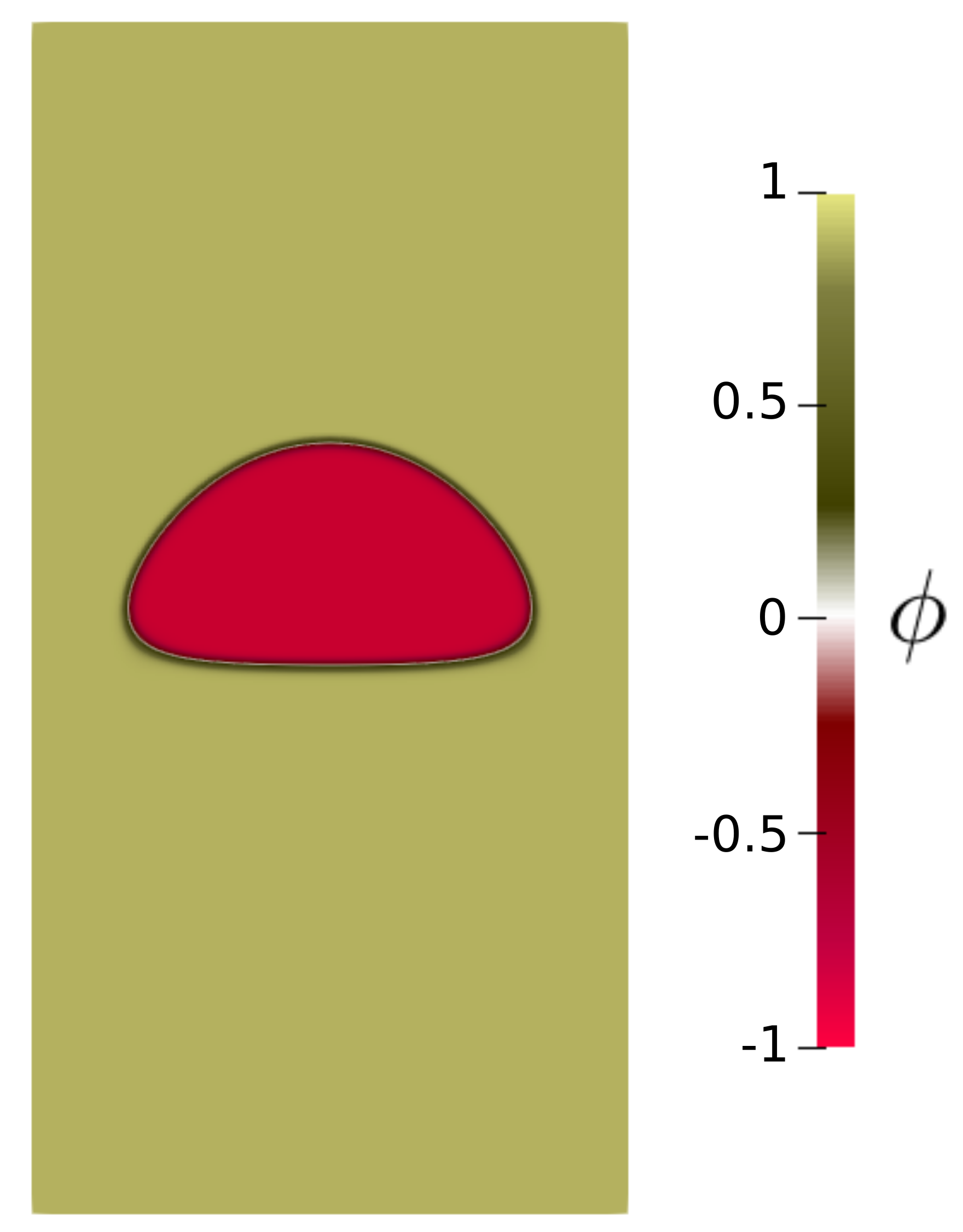}\label{mobilitytime}}
  \caption{2D rising bubble: Mobility coefficient.}
  \label{mobility}
\end{figure}

In Figure \ref{mobility}, we show the influence of the mobility coefficient for the phase-field method. We show the final contour plots of the order parameter considering a fixed mobility parameter $\gamma = 1$ and the time-dependent one with $\eta=0.05$. When $\gamma = 1$, the interface preserving capability is insufficient to keep the interface profile against the convective distortion. Therefore, at the bottom
of the bubble, the interface is subjected to a noticeable extensional distortion, which leads to an excessively low Laplace
pressure. This low Laplace pressure changes the bubble's shape, decreases the buoyancy force, and further reduces the rise velocity. On the contrary, when the time-dependent mobility model with $\eta = 0.025$ is used, the interface profile is preserved well, which gives a correct bubble shape and surface tension force calculation.

\begin{figure}[ht!]
\begin{center}
    \includegraphics[width=0.5\textwidth]{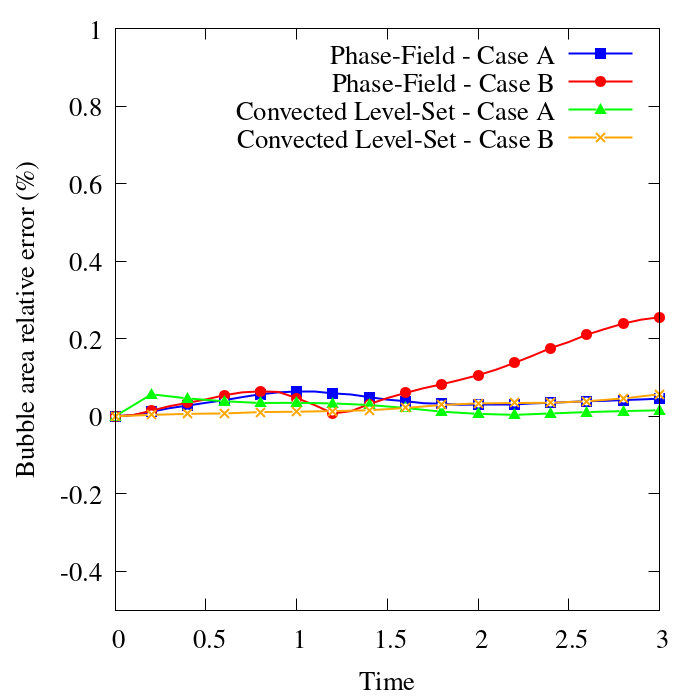}
  \caption{2D rising bubble: Mass conservation.}
    \label{mass_conservation}
  \end{center}
\end{figure}

Finally, we assess the mass conservation of each method. In Figure \ref{mass_conservation} we show the relative error between the initial and time-dependent bubble's area for both methods. The convected level-set method enforces mass conservation by a global procedure. On the other hand, the Allen-Cahn phase-field method uses a Lagrange multiplier to enforce mass conservation. Both methods preserve mass quite well (the relative error is less than 1\%) during almost the whole simulation. However, the Allen-Cahn phase-field method has issues as soon as we see filaments in the bubble. 

\subsubsection{3D rising bubble}

The bubble benchmark presented above is an excellent test to validate the methods because several results are available. However, modeling bubbles in 2D is not the best alternative since the surface tension acts in all directions, and this force significantly influences the bubble shape and the rise velocity.

We simulate a 3D benchmark proposed in \cite{adelsberger20143d} in which the results of several flow solvers are available for comparison. The only difference introduced for the 2D benchmark (aside from the additional spatial dimension) is the type of boundary condition applied at the vertical walls, which in 2D is a free-slip boundary condition, but in the 3D benchmark, it is a no-slip boundary condition. We use the same physical parameters presented in Table \ref{tab:bubble}, except for the grid size and time step.

For the convected level-set simulation we define $\lambda= 1$ and $E = 0.0125$. For the phase-field simulation we define $\epsilon = 0.0125$ and $\eta = 0.05$.
We use an adapted mesh, initially with 10 $\times$ 20 $\times$ 10 trilinear hexahedral elements, and after the refinement, the smallest element has a size of 0.00625 m. We refine the initial region where the bubble is located into four levels. The adaptive mesh refinement is based on the flux jump of the phase parameter. We apply the adaptive mesh refinement every 8 time steps.
The time-step is 0.0025 seconds.

\begin{figure}[ht!]
  \centering
  \subfloat[Allen-Cahn phase-field]{\includegraphics[width=0.35\textwidth]{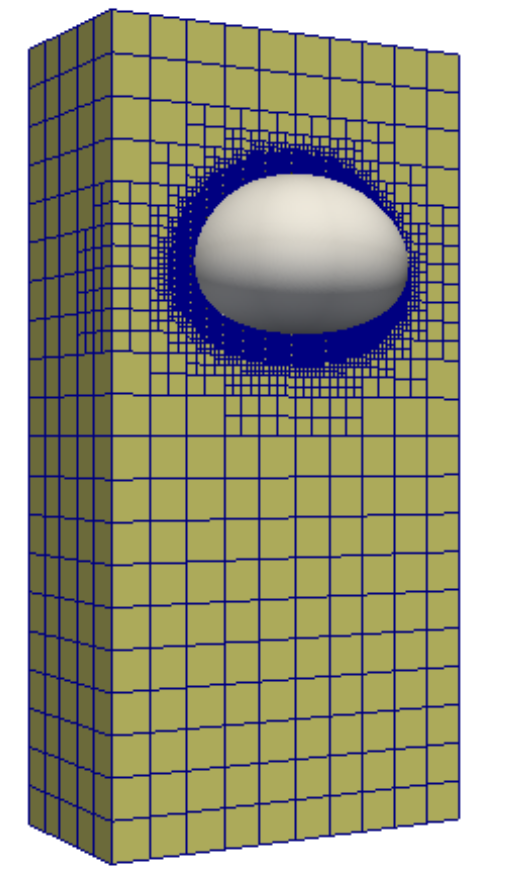}\label{case1pf}}
  \hfill
  \subfloat[Convected level-set]{\includegraphics[width=0.35\textwidth]{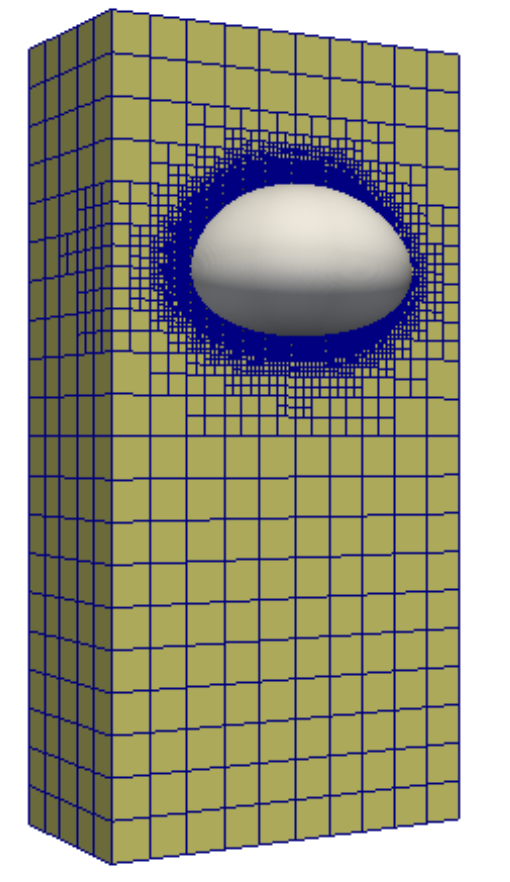}\label{case1ls}}
  \caption{3D rising bubble: Case 1 at t=3.}
  \label{3dbubble1}
\end{figure}

\begin{figure}[ht!]
  \centering
  \subfloat[Allen-Cahn phase-field]{\includegraphics[width=0.35\textwidth]{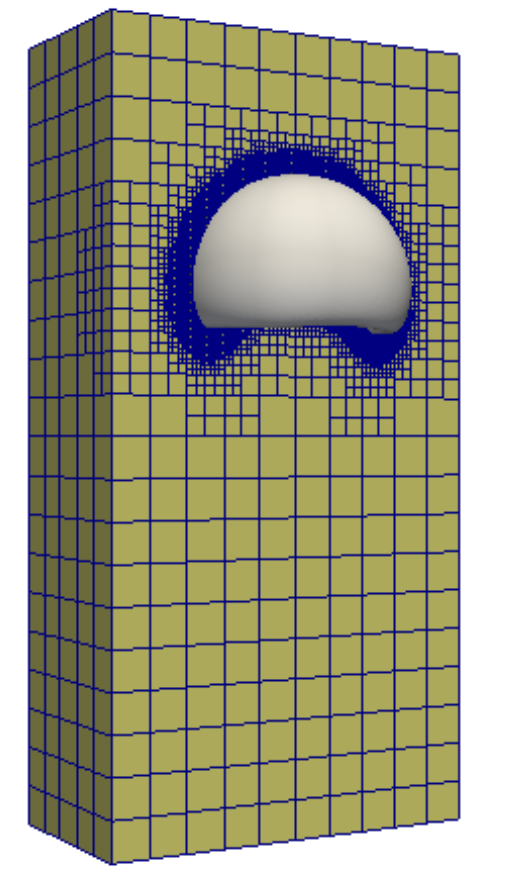}\label{case2pf}}
  \hfill
  \subfloat[Convected level-set]{\includegraphics[width=0.35\textwidth]{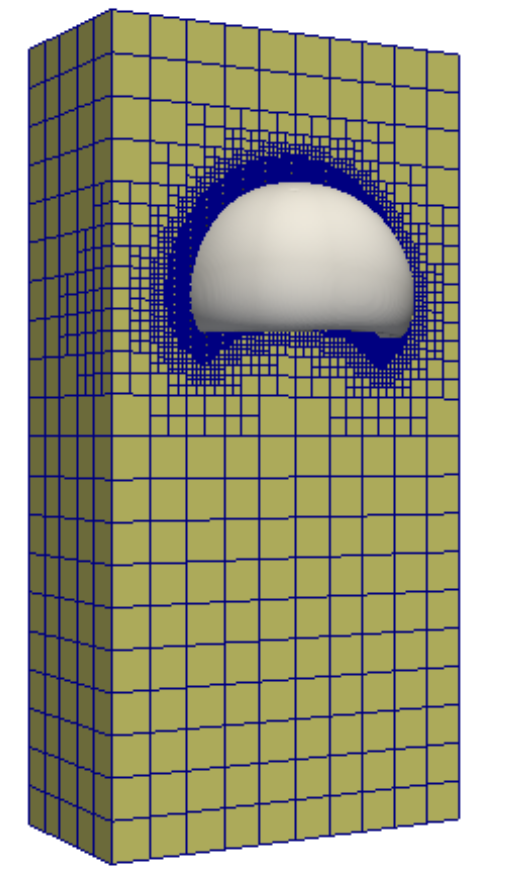}\label{case2ls}}
  \caption{3D rising bubble: Case 2 at t=3.}
  \label{3dbubble2}
\end{figure}

\begin{figure}[!ht]
  \centering
  \subfloat[Bubble shape at $t=3$.]{\includegraphics[width=0.5\textwidth]{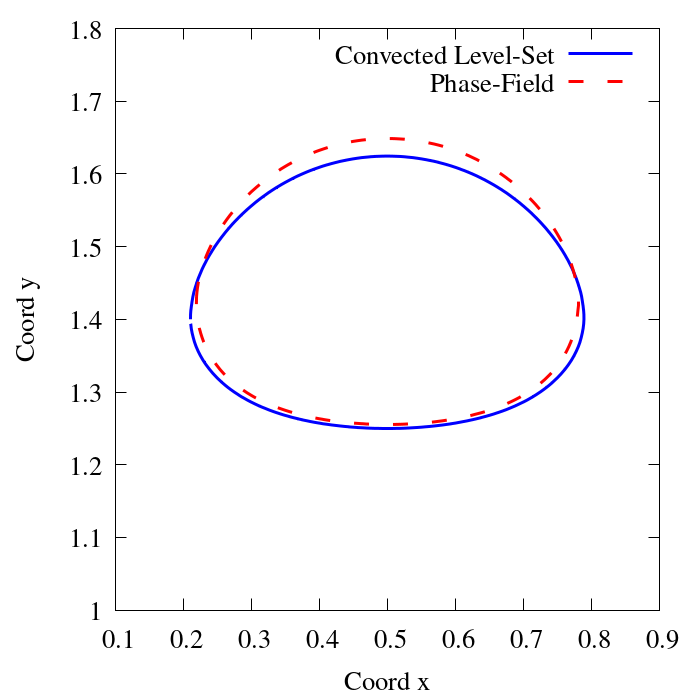}\label{benchmark13d}}
  \hfill
  \subfloat[Sphericity.]{\includegraphics[width=0.5\textwidth]{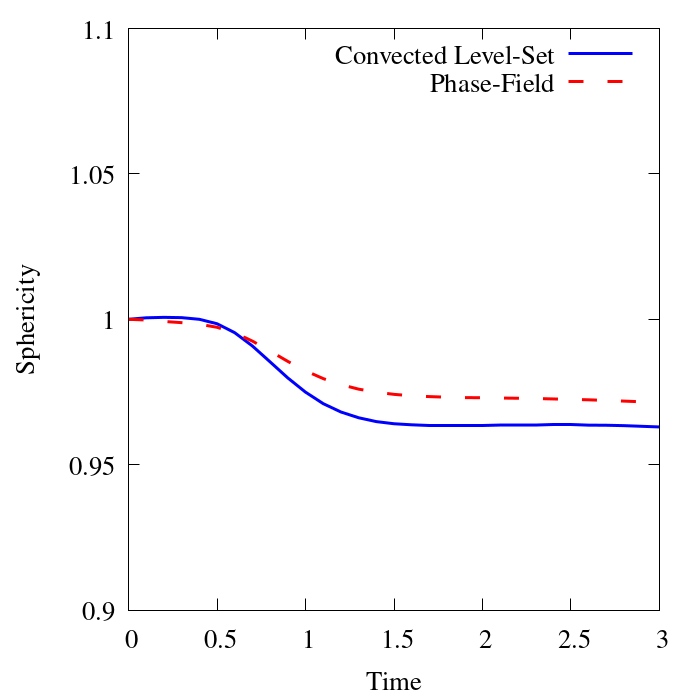}\label{circularity13d}}
    \hfill
  \subfloat[Center of mass.]{\includegraphics[width=0.5\textwidth]{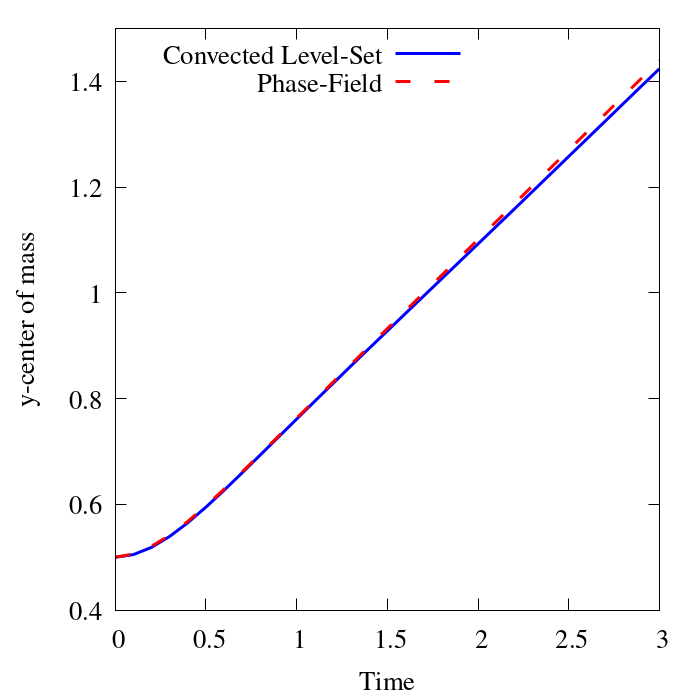}\label{mass13d}}
    \hfill
  \subfloat[Rise velocity.]{\includegraphics[width=0.5\textwidth]{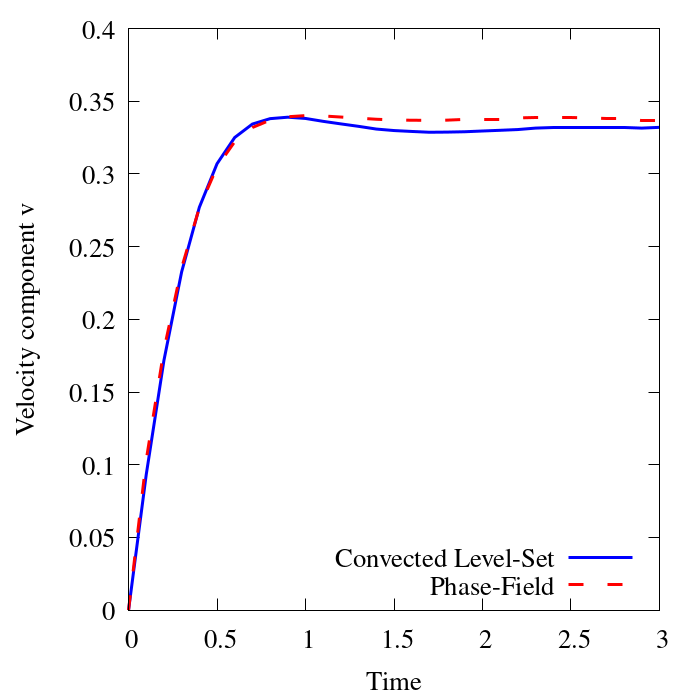}\label{risevel13d}}
  \caption{3D rising bubble: Test A results.}
    \label{qoi_case1}
\end{figure}

\begin{figure}[!ht]
  \centering
  \subfloat[Bubble shape at $t=3$.]{\includegraphics[width=0.5\textwidth]{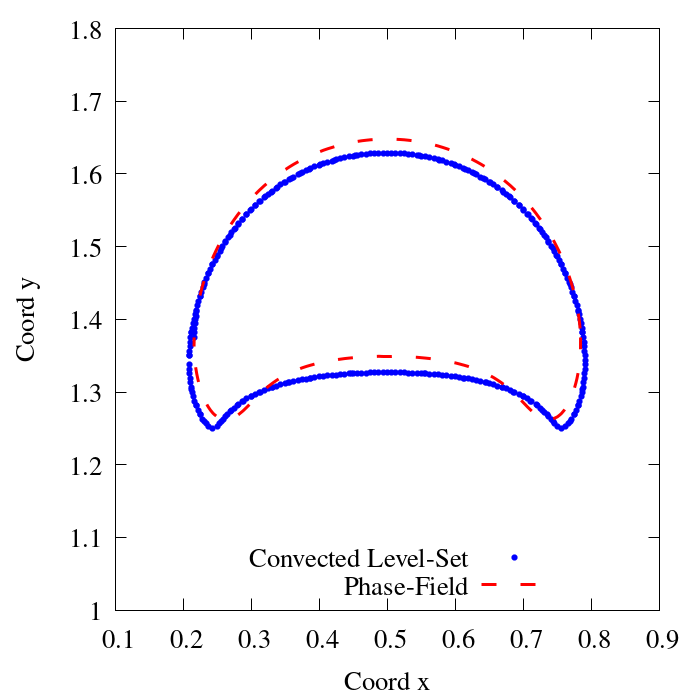}\label{benchmark23d}}
  \hfill
  \subfloat[Sphericity.]{\includegraphics[width=0.5\textwidth]{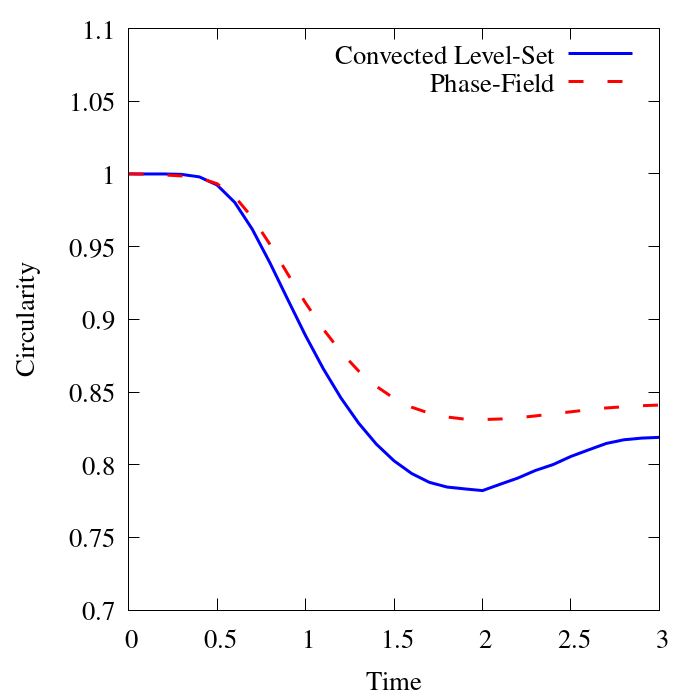}\label{circularity23d}}
    \hfill
  \subfloat[Center of mass.]{\includegraphics[width=0.5\textwidth]{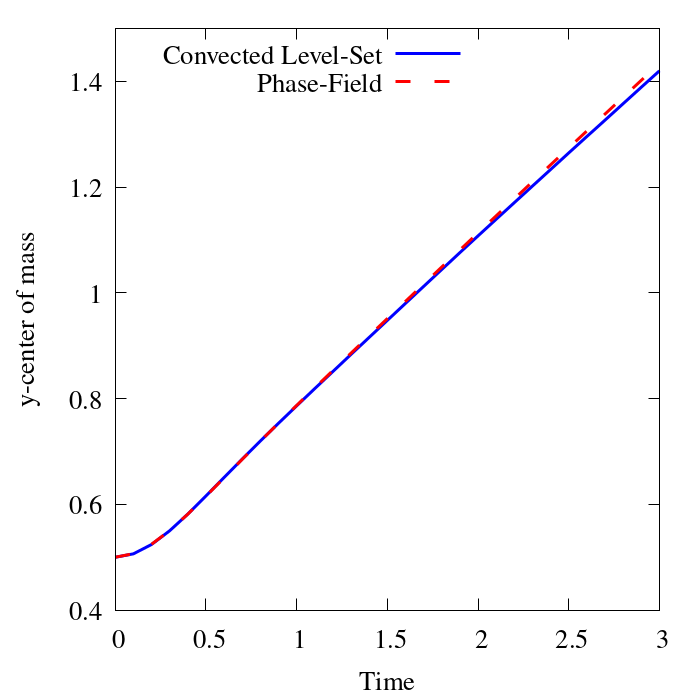}\label{mass23d}}
    \hfill
  \subfloat[Rise velocity.]{\includegraphics[width=0.5\textwidth]{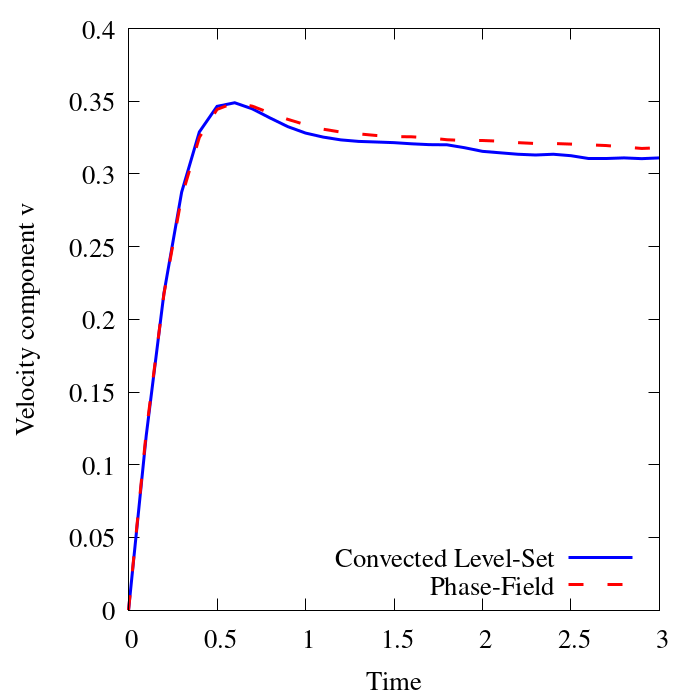}\label{risevel23d}}
  \caption{3D rising bubble: Test B results.}
  \label{qoi_case2}
\end{figure}

\begin{figure}[ht!]
\begin{center}
    \includegraphics[width=0.5\textwidth]{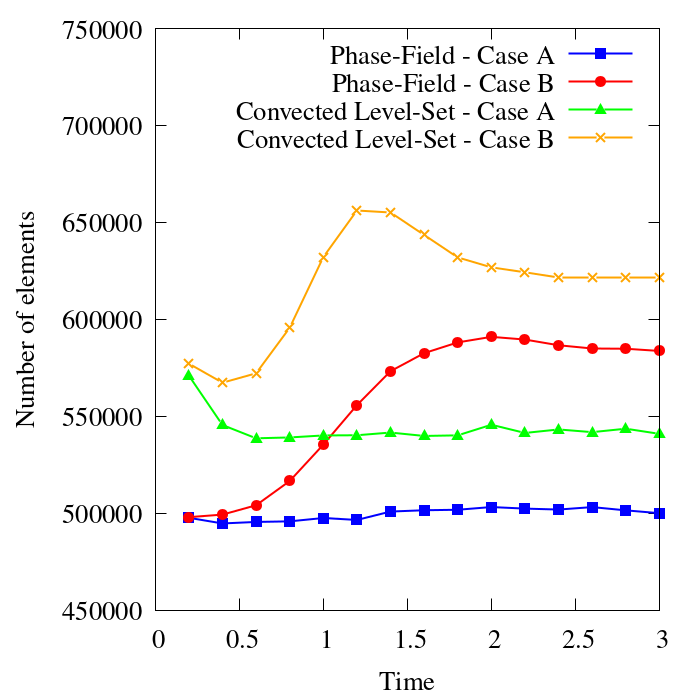}
  \caption{3D rising bubble: Number of elements.}
    \label{number_of_elements}
  \end{center}
\end{figure}

In Figures \ref{3dbubble1} and \ref{3dbubble2}, we compare the final bubble shape at t=3. for both methods. In Figures \ref{qoi_case1} and \ref{qoi_case2} we compare the quantities of interest. Although there are minor differences between the results, the convected level-set has already been verified against other solvers in \cite{Grave_Camata_Coutinho_2020}, and we may infer that, once again, the Allen-Cahn phase-field method requires more resolution than the level-set to capture the dynamics properly. Nevertheless, the phase-field has advantages against the convected level-set method, as we can see in Table \ref{tab:costs} when we compare AMR simulations with the same set-up. We show the numerical costs considering linear and nonlinear tolerances equal to $10^{-5}$. Our simulations run on the HPC cluster \textit{Lobo Carneiro}, featuring 504 CPUs Intel Xeon E5-2670v3 (Haswell) 2.3 GHz: 6048 cores and total memory of 16 TB. %Since the transition between phases is smoother in the phase-field than in the convected level-set, the mesh refinement requires few elements, and, consequently, the costs are low.
In Figure \ref{number_of_elements} we show the time history for the number of elements in each simulation, where we can see that the Allen-Cahn phase-field solutions require fewer elements in the adapted meshes than the convected level-set solutions for both cases.
Furthermore, the convected level-set uses a discontinuity-capturing operator added to the SUPG finite element formulation, which is necessary to avoid instabilities but makes the convergence of the linear and nonlinear iterations more difficult. The phase-field method does not need a discontinuity-capturing operator. Consequently, the interface movement iterations converge pretty fast. Besides, since viscosity, density, and surface tension are defined in a smoother way for the Allen-Cahn phase-field than the convected level-set, the flow's convergence is faster.

%This probably comes from the fact that the refinement is done only in function of the order parameter flux jump error, without considering the error that could appear from the velocities. On the other hand, the convected level-set represents the velocity field with less error, since it is more refined, and, consequently has a better convergence of the iterations.

\begin{table}[ht!]
  \begin{center}
  \caption{3D rising bubble: Computational costs for both methods considering the same input parameters. Comparison between the Allen-Cahn phase-field and the convected level-set.}
   \label{tab:costs}
    \begin{tabular}{|c | c c | c c|} % <-- 
      \hline
Method &  \multicolumn{2}{c|}{Phase-field} & \multicolumn{2}{c|}{Convected level-set}  \\

& Case 1 & Case 2 & Case 1 & Case 2 \\
\hline
Simulation duration (s) & 76635 & 83905 & 94223   & 99767    \\
Flow Non-Linear iterations &  7919  & 7897 & 7920 & 7911\\
Flow Linear iterations & 131736 & 128798 & 141231 & 162468   \\
Interface movement Non-Linear iter. & 4025 & 4963 & 6474 & 6424  \\
Interface movement Linear iter. & 10934 & 9914 & 72555 &  53943  \\
     \hline
    \end{tabular}
  \end{center}
  \end{table}

\subsection{Droplet impact on a thin film}\label{description}

\begin{figure}[ht!]
\begin{center}
\includegraphics[width=0.4\textwidth]{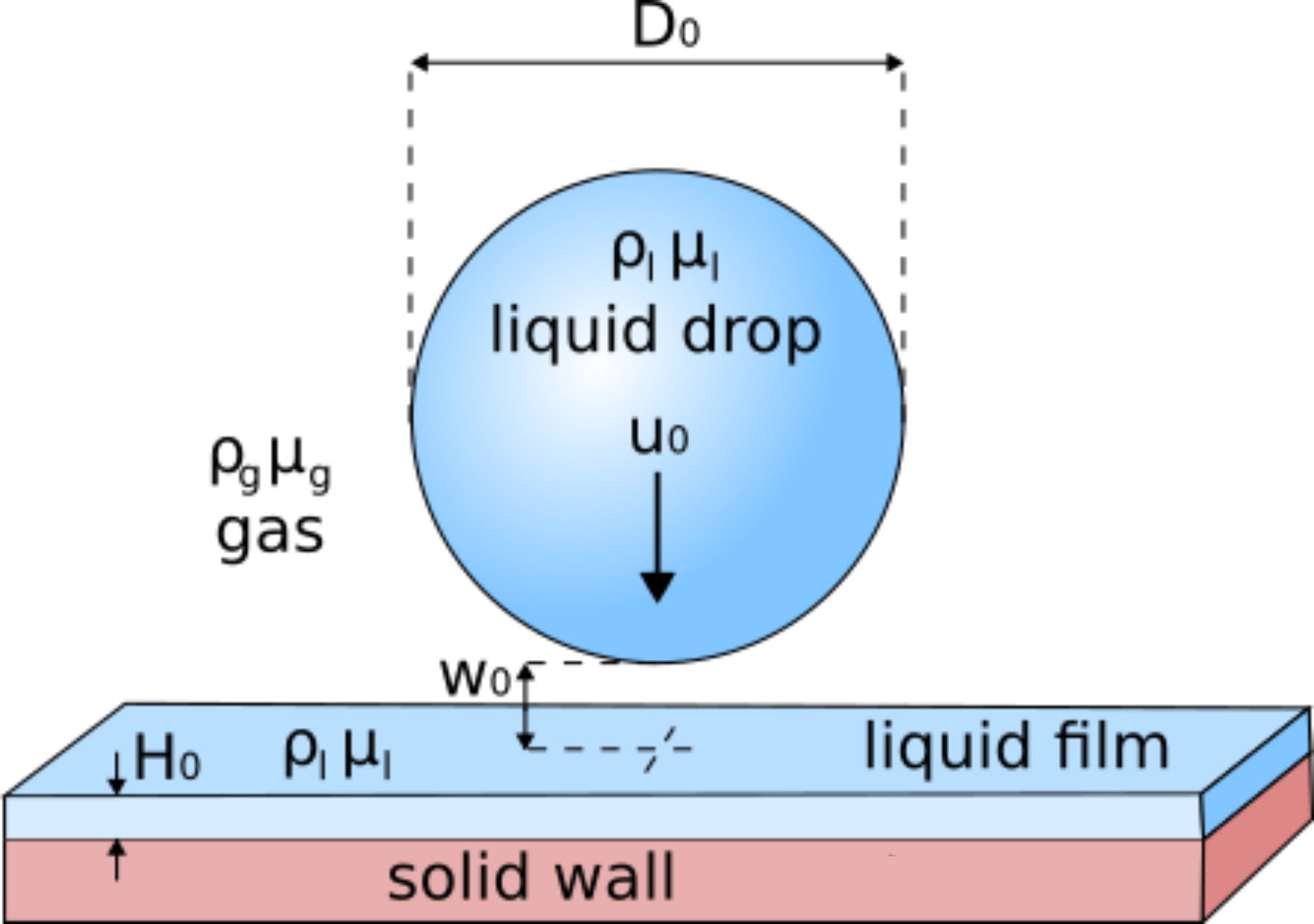}
\caption{Droplet impact: Schematic sketch.}
\label{dropscheme}
\end{center}
\end{figure}

Finally, we show the results of simulation of droplets impact. Figure \ref{dropscheme} depicts a sketch showing the main parameters of the problem of a droplet impact on a plane wall with a thin liquid film.

We consider that a liquid drop of diameter $D_0$, density $\rho_l$, and dynamic viscosity $\mu_l$ normally impacts onto a wall with a thin liquid film of the same liquid, with an impact velocity $u_0$. The thickness of the film is $H_0$. The surrounding gas has density $\rho_g$ and viscosity $\mu_g$. The liquid–gas surface tension is $\sigma_{st}$. The drop is initially placed at a distance $z_0=0.5D_0+H_0+w$ from the wall; here, the interface width $w_0=0$. At that stage of drop splashing, gravity effects are typically not important \cite{yarin2006drop,coppola2011insights}. Therefore, we neglect gravity in the present study. The computational domain is $l_x \times l_y \times l_z$, where $l_x$ and $l_y$ are the lateral dimensions of the domain in the horizontal direction, and $l_z$ is the dimension in the vertical direction. The no-slip boundary condition is used at $z=0$, while pressure boundary conditions are applied at $z=l_z$. Open boundary conditions are used at the rest of the boundaries. The initial velocity field is,

\begin{equation}
\begin{split}
&\mathbf{u} = \left\{\begin{array}{ccc}
-u_0\mathbf{k}, \textnormal{ } (\rho = \rho_l, z>0)
\\
u_w\mathbf{i}, \textnormal{ } (\rho = \rho_l, z\leq 0)
\\
0, \textnormal{ } (\rho = \rho_g)
\end{array}\right.
\end{split}
\label{uinitial}
\end{equation}

 In order to compare the simulations with other authors, we define the equations and parameters in its dimensionless form,

\begin{equation}\label{momentumad}
\rho^*\frac{\partial \mathbf{u^*}}{\partial t^*} +\rho^*\mathbf{u}^* \cdotp \nabla \mathbf{u^*} + \nabla p^* - \frac{1}{Re}\nabla \cdotp (\mu^*\nabla\mathbf{u^*}) - \frac{\mathbf{F_{st}^*}}{We} = 0 \textrm{ in }\Omega \times [0,t_f]
\end{equation} 

\begin{equation}\label{massad1}
\nabla \cdotp \mathbf{u^*} = 0 \textrm{ in }\Omega \times [0,t_f].
\end{equation}

\noindent where the superscript $*$ indicates that the parameters are in the dimensionless form. The dimensionless variables and parameters are given by,

\begin{multicols}{3}
\begin{equation}
Re =  \frac{\rho_l u_0D_0}{\mu_l}
\end{equation} 
\begin{equation}
Fr = \frac{u_0}{\sqrt{||\mathbf{g}||D_0}}
\end{equation}
\begin{equation}
We = \frac{\rho_l u_0^2D_0}{\sigma_{st}}
\end{equation}
\begin{equation}\label{xstar}
\mathbf{x^*} = \frac{\mathbf{x}}{D_0}
\end{equation}
\begin{equation}
t^* = \frac{u_0}{D_0} t
\end{equation}
\begin{equation}
\mathbf{u^*} = \frac{\mathbf{u}}{u_0}
\end{equation}
\begin{equation}
\rho^* = \frac{\rho}{\rho_l}
\end{equation}
\begin{equation}
\mu^* = \frac{\mu}{\mu_l}
\end{equation}
\begin{equation} \label{fststar}
\mathbf{F_{st}}^* = \frac{\mathbf{F_{st}}}{\sigma_{st}}.
\end{equation}
\end{multicols}

\noindent  where $\sigma_{st}$ is the surface tension coefficient.

\subsubsection{2D simulation of droplet impact on a stationary liquid film}

We now simulate a droplet impact on a stationary liquid film with large density ratio and high Reynolds numbers in a 2D domain. The liquid film height is defined as $H^*=0.15$, the diameter $D_0^*=1$ and the impact velocity $u_0^*=1$. The dimensions of the computational domain are $[10 \times 2.5]$. We set $\rho_l/\rho_g = 10^3$ and $\mu_l/\mu_g = 10^2$. Three simulations at different Reynolds numbers have been carried out ($Re=20$, $Re=100$ and $Re=1000$). The Weber number is set as $We=2000$  in all simulations.

Regarding the interface motion parameters, for the convected level-set simulation we define $\lambda = 1$ and $E=0.025$ and for the phase-field we define $\epsilon = 0.0125$ and $\eta = 0.025$. We use an adapted mesh, initially with 100 $\times$ 25 bilinear quadrilateral elements, and after the refinement, the smallest element has a size of 0.00625. We refine initially the region with a sharp phase gradient in four levels. The adaptive mesh refinement is based on the flux jump of the phase parameter error. We apply the adaptive mesh refinement every 10 time-steps and the $\Delta t = 0.001$. 

\begin{figure}[ht!]
\begin{center}
\includegraphics[width=0.6\textwidth]{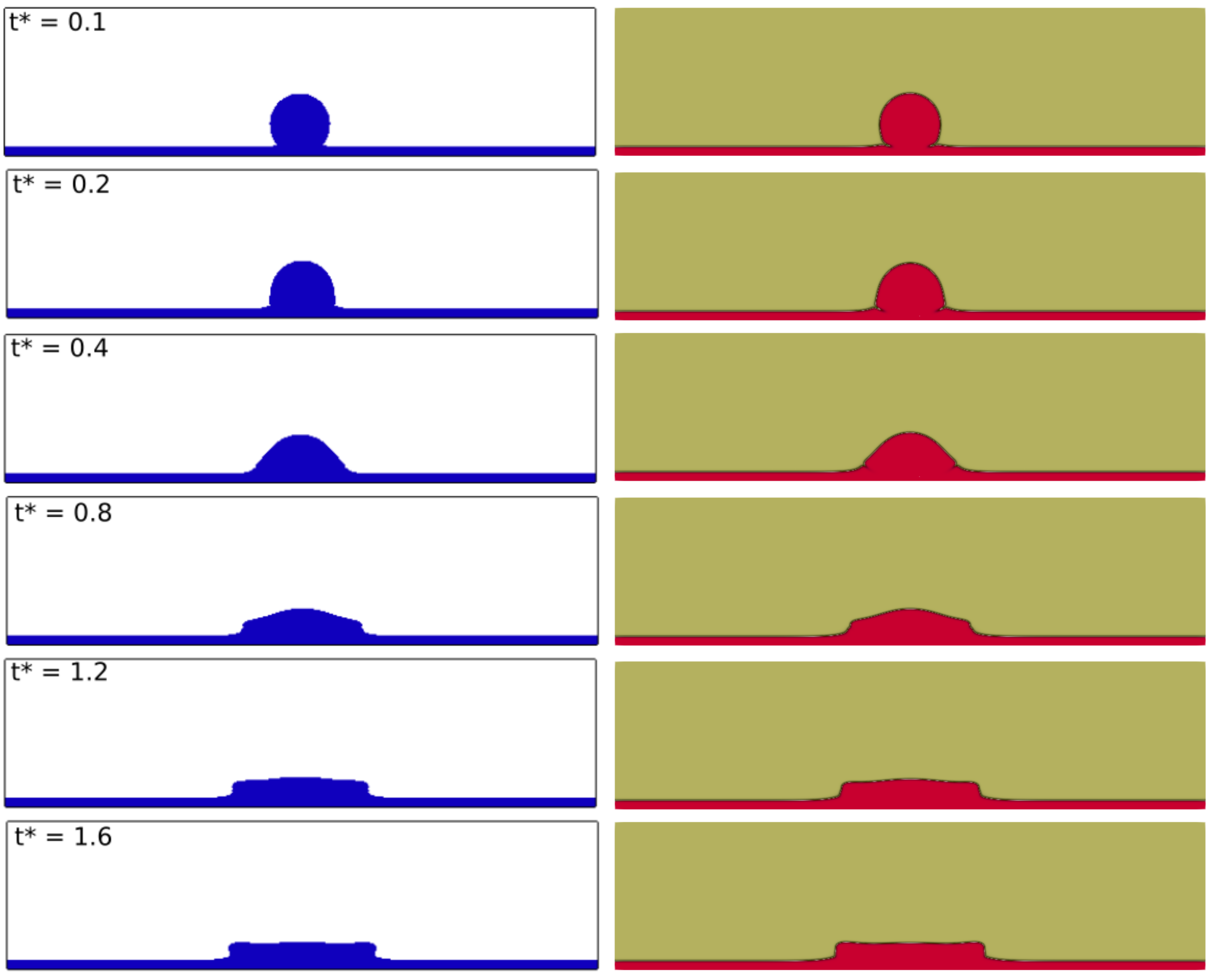}
\caption{Time evolution of droplet splashing on a thin film at Re = 20, We = 2000. Left: Convected level-set. Right: phase-field.}
\label{Re20}
\end{center}
\end{figure}

\begin{figure}[ht!]
\begin{center}
\includegraphics[width=0.6\textwidth]{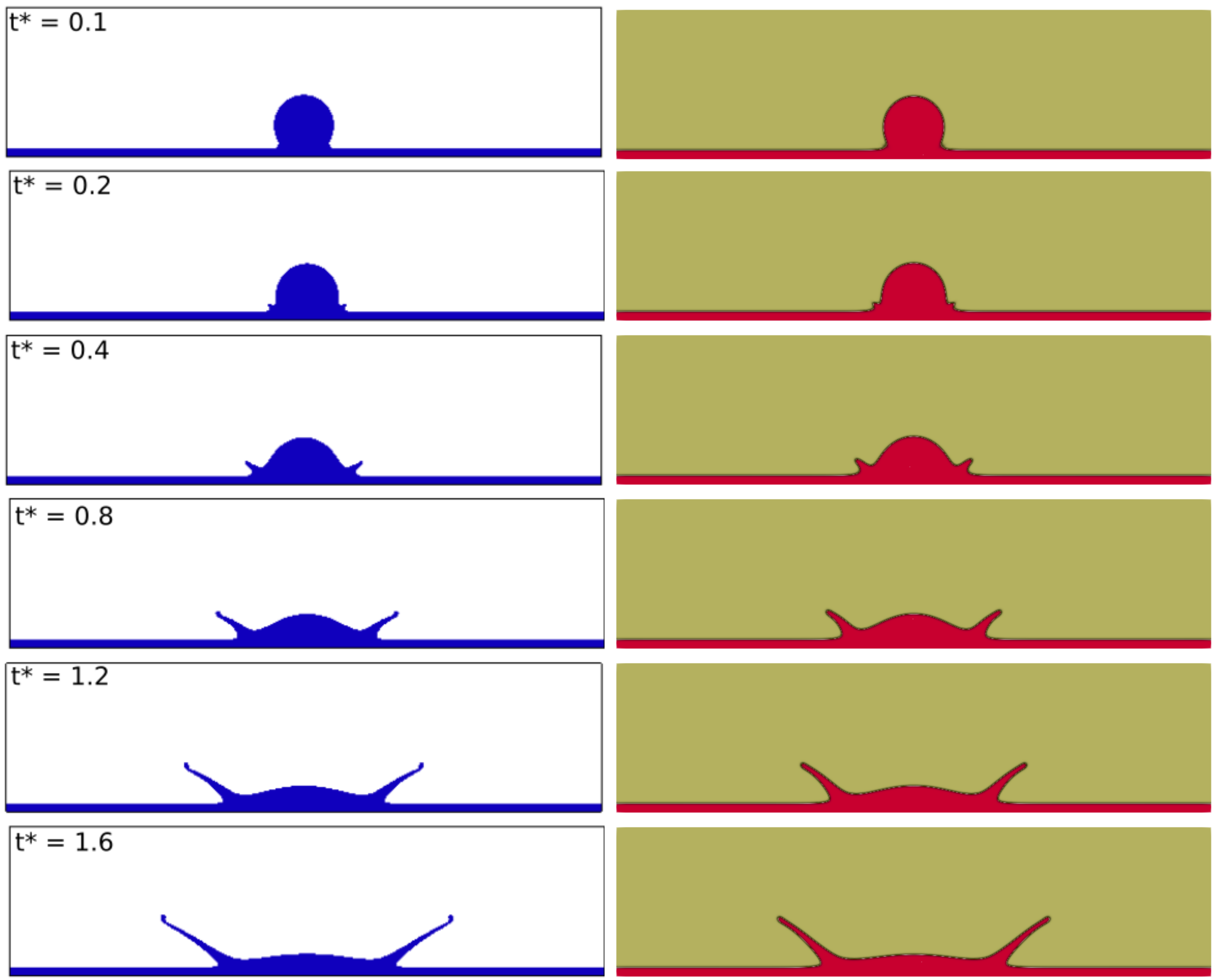}
\caption{Time evolution of droplet splashing on a thin film at Re = 100, We = 2000. Left: Convected level-set. Right: phase-field.}
\label{Re100}
\end{center}
\end{figure}

\begin{figure}[ht!]
\begin{center}
\includegraphics[width=0.6\textwidth]{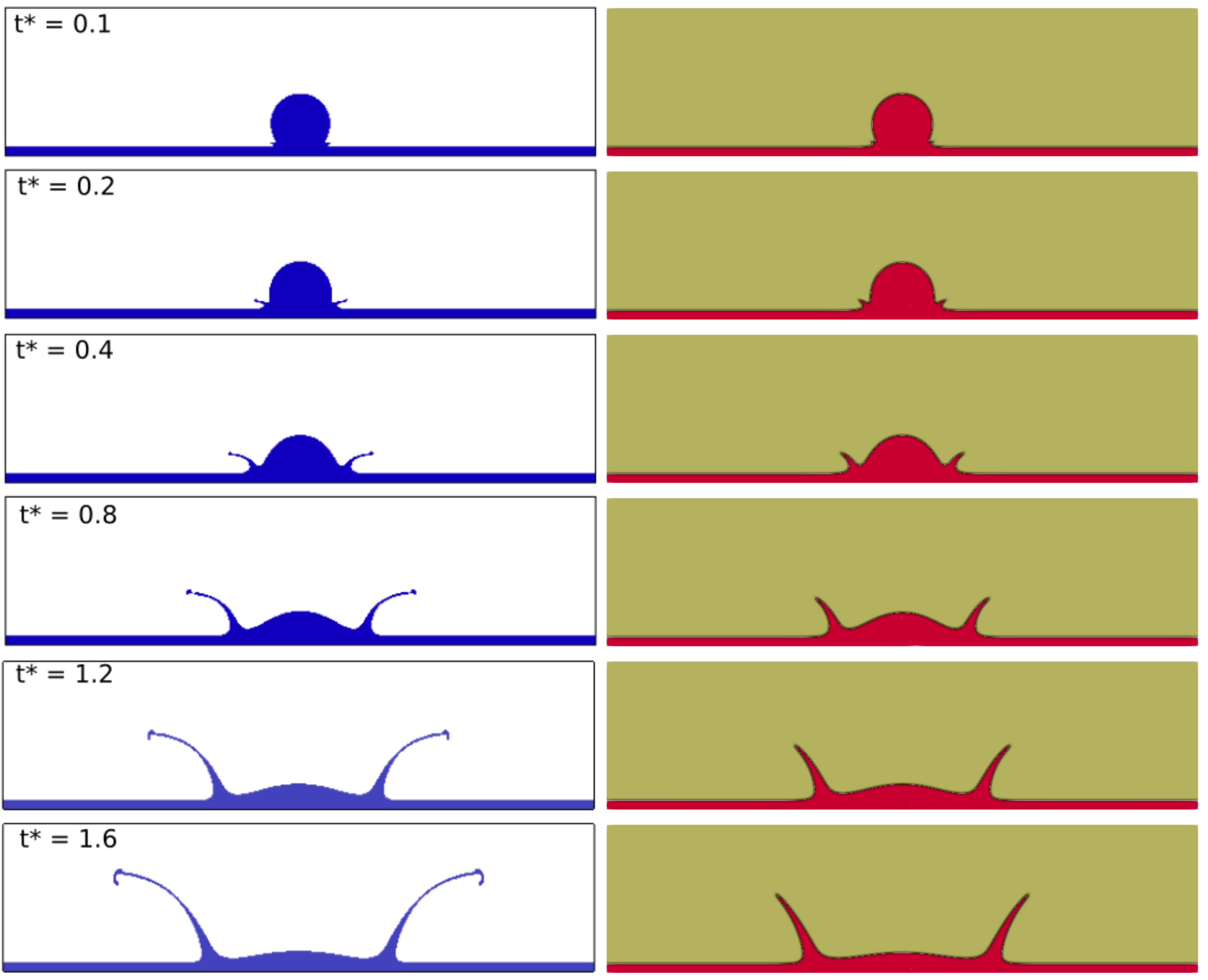}
\caption{Time evolution of droplet splashing on a thin film at Re = 1000, We = 2000. Left: Convected level-set. Right: phase-field.}
\label{Re1000}
\end{center}
\end{figure}

Figures \ref{Re20}, \ref{Re100} e \ref{Re1000} show the simulations snapshots at different time-steps. Similar dynamics were reported in the work of \cite{wang2015multiphase} and \cite{bahbah2019conservative}. However, it is possible to see that for higher Reynolds numbers, in comparison with the references, the Allen-Cahn phase-field method has a smoother solution when using the same level of mesh refinement as the convected level-set method. It is possible to improve the simulation with a refined mesh, subject to a higher computational cost.

\subsubsection{3D simulation of droplet impact on a stationary liquid film}

To further compare the methods, we simulate a 3D case of a drop splashing on a stationary thin liquid film. We consider $We=300$, $Re=2000$, $H^*=0.15$, $D_0^* = 1$ and $u_0^*=1$. The dimensions of the computational domain are $[6 \times 6 \times 3]$. We set $\rho_l/\rho_g = 100$ and $\mu_l/\mu_g = 50$.  The no-slip boundary condition is used at $z=0$, while pressure boundary conditions are applied at $z=3$. The periodic boundary conditions are used at the lateral boundaries.

Regarding the interface motion parameters, for the convected level-set simulation we define $\lambda = 1$ and $E=0.2$ and for the Allen-Cahn phase-field we define $\epsilon = 0.1$ and $\eta = 0.1$. We use an adapted mesh, initially with 30 $\times$ 30 $\times$ 15 trilinear hexahedral elements, and after the refinement, the smallest element has a size of 0.1. We refine initially the region with a sharp phase gradient in one level. The adaptive mesh refinement is based on the flux jump of the phase parameter error. We apply the adaptive mesh refinement every 10 time-steps and the $\Delta t = 0.001$. 

\begin{figure}[ht!]
\begin{center}
\includegraphics[width=0.7\textwidth]{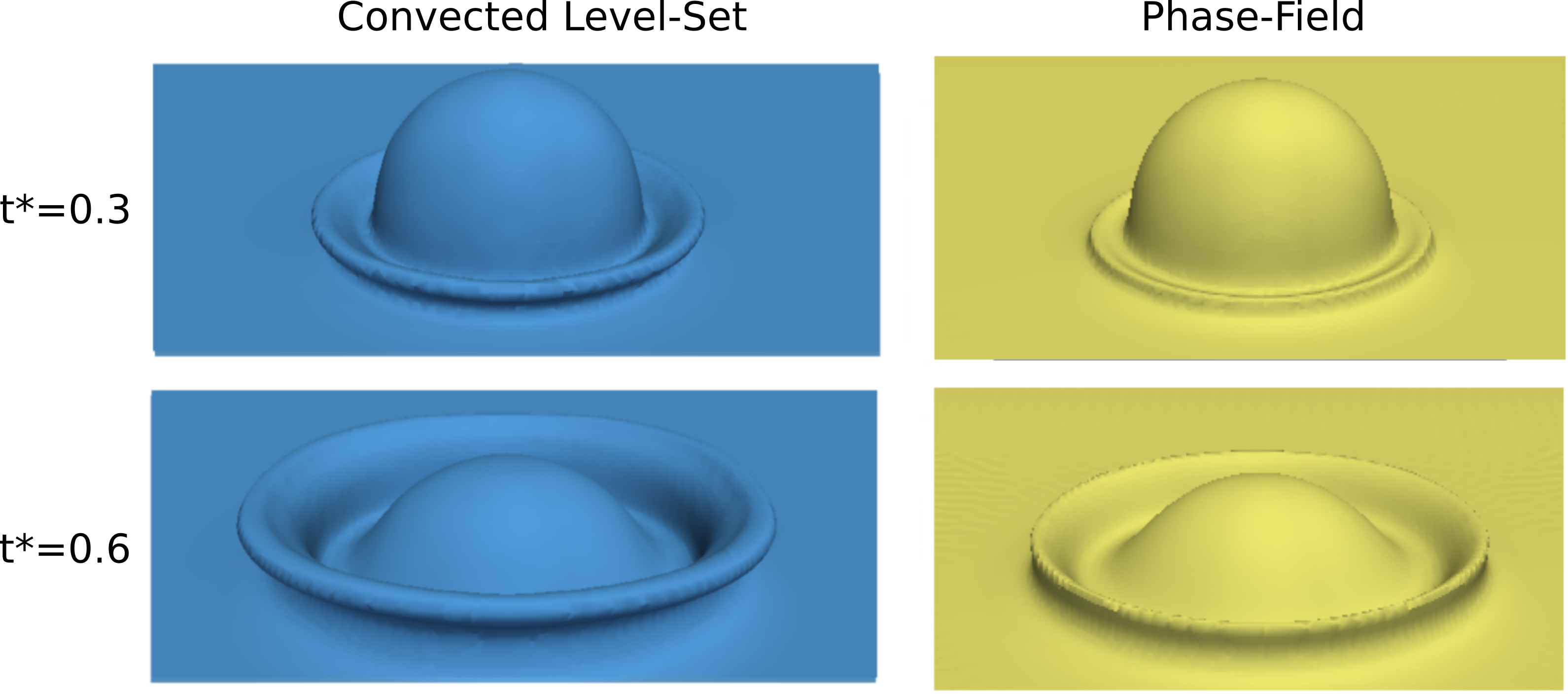}
\caption{Comparison of the snapshots of a 3D droplet splashing on a stationary thin film. Left: Convected level-set. Right: phase-field.}
\label{validade3d}
\end{center}
\end{figure}

Figure \ref{validade3d} shows two snapshots of both results, and it is possible to see that the behavior is quite similar.%, even though, once again, the phase-field method seems to need a more refined mesh to reproduce the physics correctly.
The results are very similar to the ones found in \cite{ming2014lattice}, which used a mesh five times more refined than our smallest element.

\section{Conclusions}

In this work, we have compared the convected level-set method with an Allen-Cahn phase-field method. Both methods were implemented in \texttt{libMesh}, using the same finite element formulations, time-marching schemes, solvers, and
mesh adaptivity strategies. The convected level-set implementation was already validated in \cite{Grave_Camata_Coutinho_2020}, while the Allen-Cahn phase-field validation was presented in this work.

We have run a well-known rising bubble benchmark and a droplet impact in 2D and 3D for both methods. We have seen that the physics of the bubble dynamics is well represented in both methods. However, we observe a better agreement of the convected level-set with the literature on all simulations. However, even with a coarse mesh, the Allen-Cahn phase-field method returns excellent approximations. Given the same adaptive mesh refinement and coarsening parameters, our results suggest that the Allen-Cahn phase-field method needs a more refined mesh than the convected level-set method to capture the dynamics precisely. Our observation is in line with the findings in  \cite{hua2014level}.

The Allen-Cahn phase-field method has the advantage of being stable and conserves mass with a simpler set of equations than the convected level-set. On the other hand,  the convected level-set needs an extra step to guarantee mass conservation, uses coordinate changes to evaluate Heaviside and Dirac functions, and needs a discontinuity-capturing operator to guarantee the stability of the order parameter. Nevertheless, the computational costs of both methods for the same parameters are similar, with a slight advantage for the Allen-Cahn phase-field method. 

While our results are valuable, the presented study has several limitations. Firstly, we have considered only the flux-jump of the order parameter as the error estimator to drive the mesh adaptivity. We also limited the number of refinement/coarsening steps. We know that the flow variables, other error estimators, and mesh adaption parameters influence the final solutions. Besides, our results are restricted to isotropic h-subdivision with low-order quadrilaterals and hexahedra with hanging nodes. Other strategies (e.g., p-, hp-refinement) although supported by \texttt{libMesh} were intentionally left out to simplify the comparison. Although relevant in practice, our choices are not exhaustive, and further studies are necessary. Furthermore, the role of the parameters $\lambda$ in the convected level-set method and $\eta$ in the Allen-Cahn phase-field method also deserves attention.

\section*{Acknowledgements}

This study was financed in part by the Coordenação de Aperfeiçoamento de Pessoal de Nível Superior-Brasil
(CAPES)—Finance Code 001. This work is also partially supported by CNPq, FAPERJ, ANP, and Petrobras.

%However, the convected level-set requires more iterations to calculate the interface movement because of the discontinuity-capturing operator. The number of iterations may diminish if we set the maximum number of iterations to a lower value since the residuum is already very low with one or two non-linear iterations.

%In summary, both methods demonstrate the ability to capture the interface of two-phase flows.% by having their advantages and drawbacks.

\bibliographystyle{unsrt}
\bibliography{references}  %%% Remove comment to use the external .bib file (using bibtex).
%%% and comment out the ``thebibliography'' section.

\end{document}